\documentclass[prb,aps,twocolumn,showpacs]{revtex4-1}
\usepackage{graphicx}
\usepackage{amsmath}
\usepackage[linktocpage=true,
  colorlinks=true,
  pdfborder={0 0 0},
  linkcolor=blue,
  citecolor=red,
  filecolor=yellow,
  bookmarks,
  pdftitle={Electron pairing in the presence of incipient bands in iron-based superconductors},
  pdfauthor={},
]{hyperref}
\usepackage{color}

\newcommand{\D}{\Delta}
\newcommand{\e}{\varepsilon}

\newcommand{\beq}{\begin{equation}}
\newcommand{\eeq}{\end{equation}}
\newcommand{\bea}{\begin{eqnarray}}
\newcommand{\eea}{\end{eqnarray}}

\newcommand{\bk}{{\vec k}}

\newcommand{\bq}{{\vec q}}

\newcommand{\bse}{\begin{subequations}}
\newcommand{\ese}{\end{subequations}}
\newcommand{\bwt}{\begin{widetext}}
\newcommand{\ewt}{\end{widetext}}

\newcommand{\bsu}{\begin{subequations}}
\newcommand{\esu}{\end{subequations}}

\graphicspath{{./figures/}}

\begin{document}

\title{ Electron pairing in the presence of incipient bands  in iron-based superconductors}
\author{Xiao Chen$^1$, S. Maiti$^{1,2}$, A. Linscheid$^1$, and P. J. Hirschfeld$^1$}
\affiliation {~$^1$Department of Physics, University of Florida, Gainesville, FL 32611}
\affiliation {~$^2$National High Magnetic Field Laboratory, Tallahassee, FL 32310}
\date{\today}

\begin{abstract}   Recent experiments on certain Fe-based superconductors have hinted at
a role for paired electrons  in ``incipient" bands that are close to, but do not cross the Fermi level.
Related theoretical works disagree on whether or not strong-coupling superconductivity is required to
explain such effects, and whether a critical interaction strength exists. In this work, we consider various versions of  the model problem of pairing of electrons in the presence of an incipient band, within a simple multiband weak-coupling BCS approximation.  We categorize the  problem into two cases: case(I) where superconductivity arises from the ``incipient band pairing'' alone, and case(II) where it is induced on an incipient band by pairing due to Fermi-surface based interactions.  Negative conclusions regarding the importance of incipient bands have been drawn so far largely based on case(I), but we show explicitly that models under case(II) are qualitatively different, and can explain the non-exponential suppression of $T_c$, as well as robust large gaps on an incipient band.      In the latter situation, large gaps on the incipient band do not require a critical interaction strength. We also model the interplay between phonon and spin fluctuation driven superconductivity and describe the bootstrap of electron-phonon superconductivity by spin fluctuations coupling the incipient and the regular bands.  Finally, we discuss the effect of the dimensionality of the incipient band on our results.  We argue that pairing on incipient bands may be significant and important in several Fe-based materials, including  LiFeAs, FeSe intercalates and FeSe monolayers on strontium titanate, and indeed may contribute to high critical temperatures in some cases.
\end{abstract}

\pacs{74.20.-z, 74.70.Xa}

\maketitle

\section{Introduction}

The standard paradigm for $s_{\pm}$ pairing in Fe-based superconductors(FeSC)\cite{HosonoKuroki2015,CH_PhysToday2015,HKMReview2011,ChubReview2012,WenLiReview2011} relies on the existence of a hole-like Fermi surface (FS) near $\bk=0$ and an electron-like  FS near $\bk'=(\pi,0)$ in the 1-Fe Brillouin zone, and symmetry-related points. Repulsive interband interactions and approximate nesting then lead, within this simplified picture, to a strong peak in the particle-hole susceptibility at $\bq=\bk-\bk'=(\pi,0)$, which drives a spin fluctuation pairing interaction that can condense pairs only if the superconducting (SC) order parameter changes sign between the two pockets\cite{mazin08,kuroki08}. Beginning in  2010 with the discovery of superconductivity in the alkali-intercalated FeSe materials\cite{Guo2010,Sadovskii2012,Yu2013}, this paradigm was challenged by the subsequent remarkable discovery\cite{Qian2011,Zhang2011,Fangdwave} that all hole bands in these materials, with optimal critical temperatures greater than 40K, were below the Fermi level. Several groups pointed out that repulsive interactions at the Fermi level remained among the electron Fermi surface pockets, and could lead to $d$-wave pairing with significant critical temperatures\cite{Fangdwave,Maierdwave}.

Ref. \onlinecite{Fangdwave} also pointed out that pairing in an $s_{\pm}$ channel with sign changing gap was still quite competitive, despite
the fact that the hole band was $\sim$ 90 meV below the Fermi level, indicating presence of substantial spectral weight of the spin fluctuations. While this ``incipient'' $s_{\pm}$ possibility was considered\cite{HKMReview2011}--along with the $d$-wave state and a more subtle $s$-wave state that changed sign between two hybridized electron pockets in the 2-Fe zone--as a possible candidate for pairing in these materials, it did not receive a great deal of attention. This is probably because of the general feeling in the community that incipient bands (we use this term ``incipient'' here to mean bands away from the Fermi level, but within a `pairing' cutoff energy) do not play an important role in superconductivity. As discussed in Ref. \onlinecite{HKMReview2011}, in a simple model
of electron pocket - hole pocket $s_{\pm}$ pairing, if the hole pocket maximum moves below the Fermi level by an energy $|E_{g}|$, the  dimensionless pairing strength ( $v$) in this channel is reduced: $v\rightarrow v^{2}\log \Lambda/|E_g|$, where $\Lambda$ is the pairing cutoff. This suggests that within weak coupling theories, one gets a strong suppression of $T_c$ as $|E_g|$ is increased.

The discussion of the role of the incipient  band in pairing in FeSCs was revived by several new experiments. The first was the discovery by angle-resolved photoemission (ARPES) that the electronic dispersion in FeSe monolayers on SrTiO$_{3}$ (STO), with extremely high critical temperatures of around 70K (ARPES gap closing)\cite{XJZhou2013} was similar to the alkali-intercalated FeSe systems, namely the central hole pocket was pushed below the Fermi surface by   $\sim$ 80 meV. The second was the observation by  Miao et al. \cite{HongDingLiFeAs2015} of a superconducting gap on one of the hole bands of LiFeAs as it fell below the Fermi level with electron doping by Co. Here it was found that the gap was suppressed only rather weakly in this process, compared to one's naive expectations according to weak coupling BCS theory, and survived at least up to band extremum values of $E_{g}\sim -8$ meV.  These authors  suggested that, because the variation of the gap on the hole band was  gradual through the Lifshitz transition,  a standard weak-coupling scenario was unlikely. Finally, a more recent experiment has  reported a Fermi surface without hole pockets, very similar to the FeSe monolayers, in the new LiFeOH- intercalated FeSe material.\cite{DLFeng2015}

There have been some theoretical efforts addressing these systems and the idea of incipient band pairing. The most
 relevant work
is from Bang,\cite{Bang} who explored the evolution of $T_c$ across a Lifshitz transition of a band in a model for Ba$_{1-x}$K$_x$(FeAs)$_2$. He pointed out that $T_c$ may remain substantial in the presence of an attractive intra-band interaction.  Considering only interband interactions, Bang also concluded that the gap induced in the incipient band will be significant and should show up as a shadow gap in the ARPES spectrum. Leong and Phillips\cite{Phillips2015} recently also considered a model specific to LiFeAs within weak-coupling Eliashberg theory and argued that Coulomb interactions can stabilize a robust isotropic gap in a ``shallow" band which barely crosses the FS. Hu et al. \cite{FCZhang2015} attempted a more realistic calculation of the effect of an incipient band in the LiFeAs system, and concluded that one needs to consider large couplings in order to explain the experiments, and also a minimum pairing interaction to induce a gap on the incipient band: which is apparently contrary to the message in Bang's work.

In this work, we extend Bang's idea, within a simple multiband BCS approximation, to perform a systematic study of pairing in the incipient band in FeSCs. We clarify that there are two classes of problems that arise: (I) when pairing is driven by interactions only with the incipient band; and (II) when pairing is induced in the incipient band due to an already stabilized SC ground state due to other bands that cross the Fermi level.  We argue that, unlike the result in Hu et al., there is no minimum interaction strength except for a special instance of case(I). We show that the usual expectation of strong suppression of gaps and $T_c$ apply only to case(I). The models for case(II) suggest that  (a) the problem is well defined and can be treated in weak coupling; (b) there is no minimum interaction strength needed to induce SC;  (c) the induced gap on the incipient band is comparable to and can be larger than other gaps in the system; (d) spin fluctuations  (interband interactions)  are crucial to induce significant pairing in the incipient band; (e) spin fluctuations can bootstrap an existing phonon based interaction and yield a larger $T_c$ and a sizable SC gap; (f) the dimensionality of the incipient band can play a role in  determining  the magnitude of the effect on SC.

We arrange the article in the following way: In Sec. \ref{Sec:Model} we discuss the formulation of cases(I) and (II) and discuss the literature in some detail; In Sec. \ref{Sec:case1} we discuss case(I) where  standard results are recovered. In Sec. \ref{Sec:case2}, our main section, we discuss all aspects of  case (II) and compare with the results mentioned above. In Sec. \ref{Sec:Disc} we
 present our ideas in the context of the experimental situation vis a vis particular FeSC materials, and
summarize in Sec. \ref{Sec:Conclusion}. In the Appendix,  we summarize the effect of three dimensionality of the density of states (DOS) in the incipient hole band on the results obtained in this work.

\section{Models for SC in the incipient band}\label{Sec:Model}

\begin{figure*}[htbp]
$\begin{array}{cc}
\includegraphics[width=0.397\textwidth]{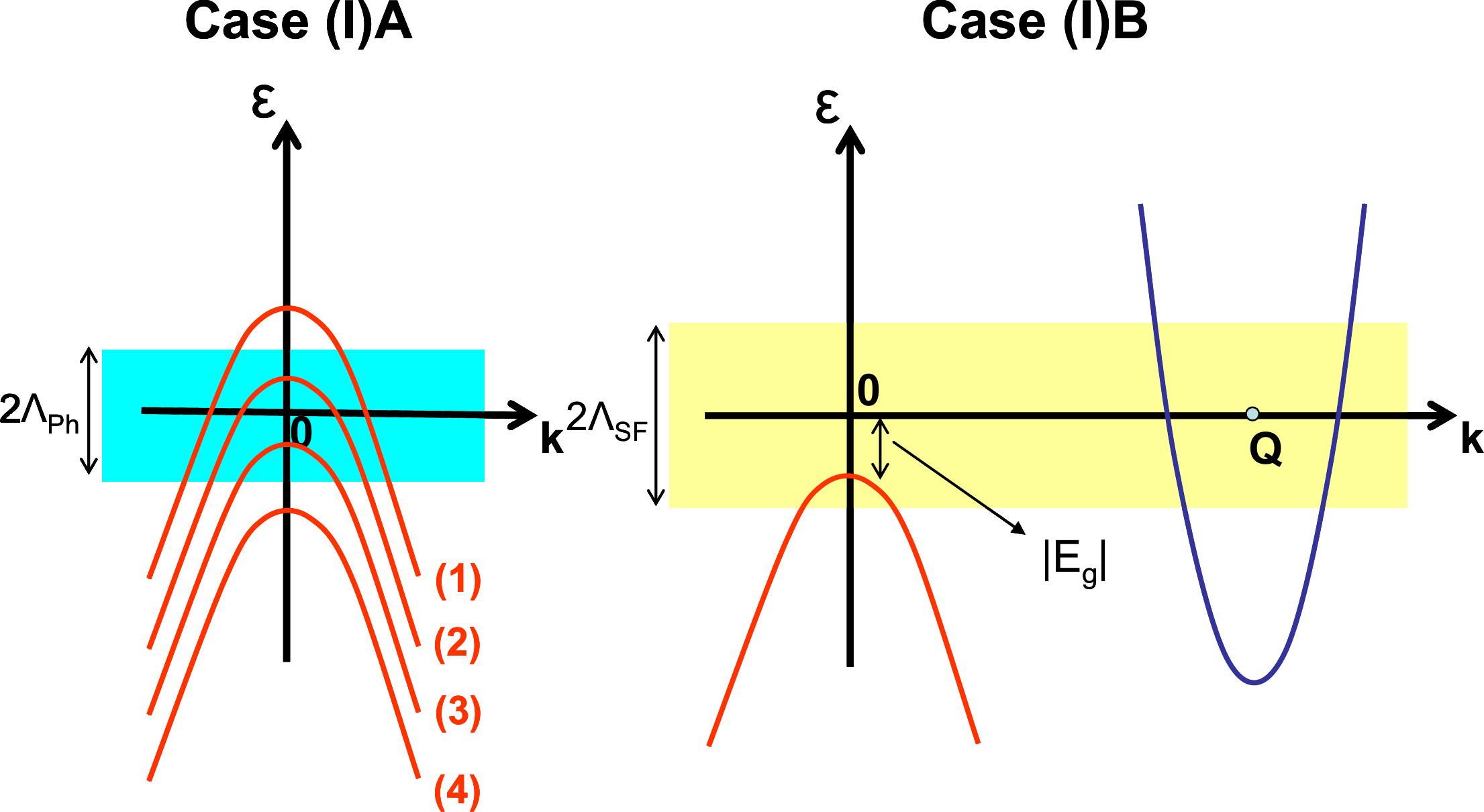}\nolinebreak
\includegraphics[width=0.603\textwidth]{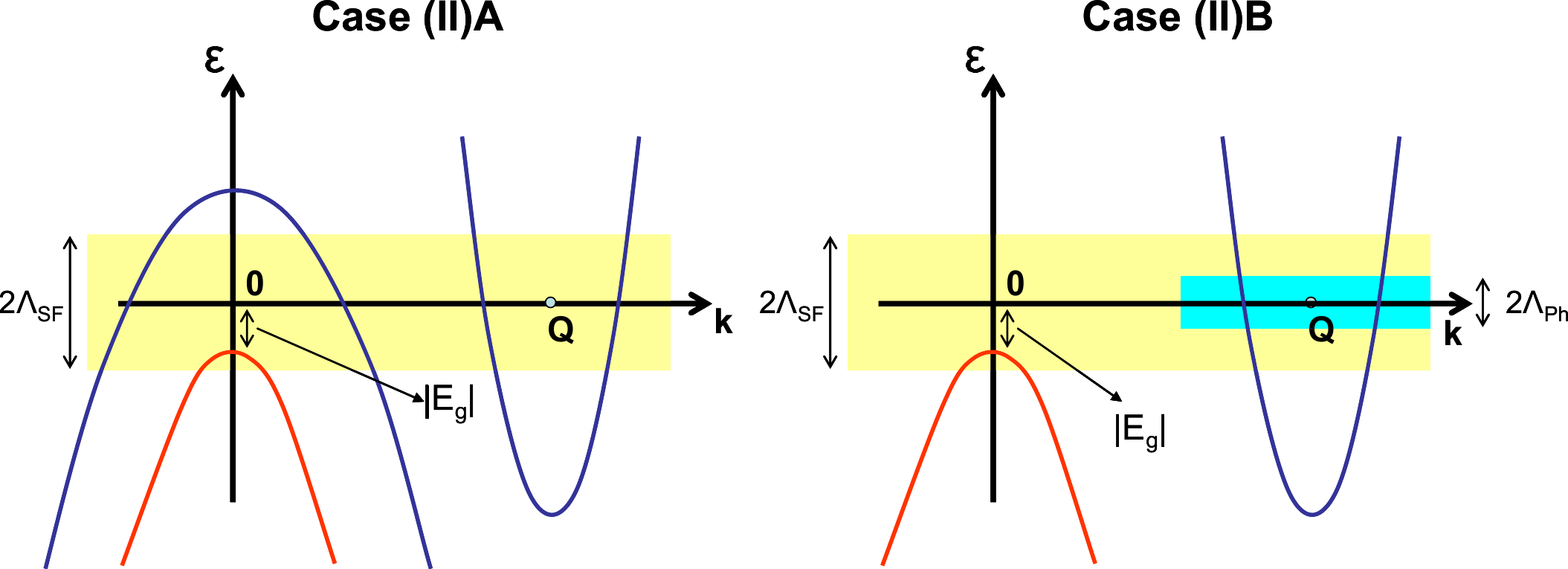}
\end{array}$\caption{\label{fig:1}(Color online) Case(I)A: The 4 instances of the hole band correspond to (1)regular band (2)shallow band (3)incipient band (4) vegetable band. This is a representation of a conventional case with phonon driven interactions with cutoff $\Lambda_{\text{ph}}$ (blue region). Case (I)B:  Representative of the incipient case for spin fluctuation driven (cutoff $\Lambda_{\text{sf}}$-yellow region) SC. Case (II)A: Representative of the incipient case for the situation where SC is driven by spin fluctuations in the regular (blue) bands. SC in the incipient band is induced by the same interaction. Case(II)B: Representative of the incipient case for the situation where SC is driven by phonons in the regular (blue) band. SC is induced in the incipient band through spin fluctuations.}
\end{figure*}

The two cases mentioned in the introduction need to be distinguished, as they give fundamentally different results.  We have sketched the various possibilities in Fig. \ref{fig:1}.   Case (I)A, which considers pairing in the incipient band when the driving pairing interaction (phonon-mediated,   i.e. attractive) involves states in the incipient band itself, is the case usually imagined when the irrelevance of incipient band pairing is claimed. Case(I)B considers spin fluctuation as the driving pairing interaction that connects a regular band and an incipient band. This was  discussed in Ref. \onlinecite{HKMReview2011}  and numerically explored by Bang\cite{Bang},  with the result that $T_c$ is drastically suppressed as the  incipient band  extremum  $|E_g|$ is increased,  unless an attractive intraband interaction is added. Our study of case(II) is also comprised of 2 categories, which we use to explore spin fluctuation driven SC and phonon-driven SC. In Case(II)A, a (repulsive) spin fluctuation mediated (pairing cutoff $\Lambda_{sf}$) SC is stabilized in the already existing bands, and the same spin fluctuations induce SC in the incipient band. In Case(II)B, an (attractive) phonon-mediated (pairing cutoff $\Lambda_{ph}$) interaction results in SC in the electron pockets and the spin fluctuations induce SC on the incipient hole band. We assume here that interband phonon coupling is weak. This will serve as our paradigm for spin fluctuations bootstrapping the electron-phonon mediated SC.

We take this opportunity to comment on some other works addressing incipient band pairing.  Miao et al. presented a curve labelled ``BCS weak coupling" which indicates a gap on the incipient band  falling rapidly compared with experiment as the band sinks below the Fermi level, without giving details of the calculation. They imply that this disagreement rules out BCS weak-coupling type physics.  Furthermore, they argue  that the large size of the gap on the incipient band rules out ``proximity-coupled" superconductivity, i.e. the possibility  that the superconductivity caused by pairing of states at the Fermi level in other bands could induce a large gap in the incipient band.

We now consider the work by Hu et.al. \cite{FCZhang2015}, where the experiment in Ref.  \onlinecite{HongDingLiFeAs2015} motivated them to study a more realistic 3-orbital model for LiFeAs with a next-nearest neighbor intersite BCS-like pairing ansatz. These authors claimed that the results of the experiment could be understood on the basis of requiring a strong pairing strength coupling the incipient band to the Fermi surface pocket and having a minimum  threshold for the pairing strength, thereby suggesting that  strong coupling physics is required. We believe that this conclusion is incorrect, and in fact will show that the strict requirement for a minimum pairing strength only arises in case(I)A, and hence is not applicable to FeSCs.   We discuss this point further below.

Another realistic model was studied in Ref. \onlinecite{Phillips2015} where a 5-band model was considered with spin fluctuation interactions scattering electrons near the Fermi surface, and   Coulomb  interactions renormalized to a low-energy cutoff scattering electrons from a shallow band to one of the other hole  bands. While these authors reported an enhanced $T_c$ and the largest gap on the shallow pocket, some  caveats remain: (1) For the tiny shallow pocket, one might expect the Coulomb repulsion to be strong within the band, yet this was dropped, retaining only a repulsive interband Coulomb interaction with the other hole band; (2), a constant DOS was assumed in the derivation of the Eliashberg equations used, despite the low energy scale of the shallow band; (3) strictly speaking, incipient band pairing was still not considered.

The work by Bang\cite{Bang} correctly captured the idea of incipient band pairing, and also pointed out that the induced gaps may be significant  in the incipient case.  However, he specialized to parameters appropriate for Ba$_{1-x}$K$_x$Fe$_2$As$_2$, where $T_c$ had been observed to vary only weakly through a Lifshitz transition\cite{Xu13}.

We keep the modelling  simple and extend Bang's  idea systematically to all the cases mentioned above, allowing us to discuss analytical results  in important cases. We study the gradual evolution of the gaps and $T_c$ for every case  and show that models representing case(II) have the potential to explain the recent experiments.

\section{SC in incipient band - Case(I)}\label{Sec:case1}

For completeness, we revisit the conventional case(I) in some detail and show, within weak coupling, how one can qualitatively reproduce the previously known results. Since this part is intended to be a demonstration of principle, we strive to keep the presentation of case (I)A   (see Fig.~\ref{fig:1}) simple. The multiband case (I)B follows from  a treatment  similar to that presented in Ref. \onlinecite{Bang}, so we do not dwell on details.
\subsection{Case (I)A}
Within the weak coupling BCS treatment of the problem, our simple 1-band example (I)A involves solving the following gap equation at temperature $T$ (with $\hbar=k_{{\rm B}}=1$  and unit volume):
\bea\label{eq:1}
\D_{\bk} &=& -\int\frac{d^2k}{(2\pi)^2} V_{\bk,\bk'} \frac{\D_{\bk'}}{2E_{\bk'}}\tanh\frac{E_{\bk'}}{2T},
\eea
where $E_{\bk}=\sqrt{\e_{\bk}^2 + \D_{\bk}^2}$. The hole band dispersion is
\beq\label{eq:2}
\e_{\bk}=-\frac{k^2}{2m} + E_g,
\eeq
where $m$ is the band mass and $E_g$ is the shift of the hole band. We work with energies relative to the chemical potential $\mu$ and hence set $\mu=0$. $E_g>\Lambda$ is the regular BCS case (instance-1), $\Lambda>E_g>0$ corresponds to the shallow band case(instance-2), $0>E_g>-\Lambda$ corresponds to the incipient case (instance-3) and $-\Lambda>E_g$ is the vegetable case (instance 4) where the band does not participate in SC. Choosing the (attractive intraband) pairing interaction $V_{\bk,\bk'}=V_{\text{ph}}<0$ for $|\e_{\bk}|,~|\e_{\bk'}|<\Lambda$, the order parameter $\D_{\bk}$ becomes a constant $\D$. After we solve these equations, we get $T_c$ as a function of $E_g$. As long as $E_g>\Lambda$, we remain in the conventional BCS regime (instance-1 in Fig. \ref{fig:1}). Interesting effects  arise when the band becomes shallow (instance-2) and incipient (instance-3). This marks the first step of departure from a conventional BCS approach because the band edge now falls within the pairing energy scale. Already at this stage we note that all the corrections to the BCS theory of $\mathcal{O}(\Lambda/E_F)$ become relevant.

\begin{figure}[htbp]
$\begin{array}{c}
\includegraphics[width=0.8\columnwidth]{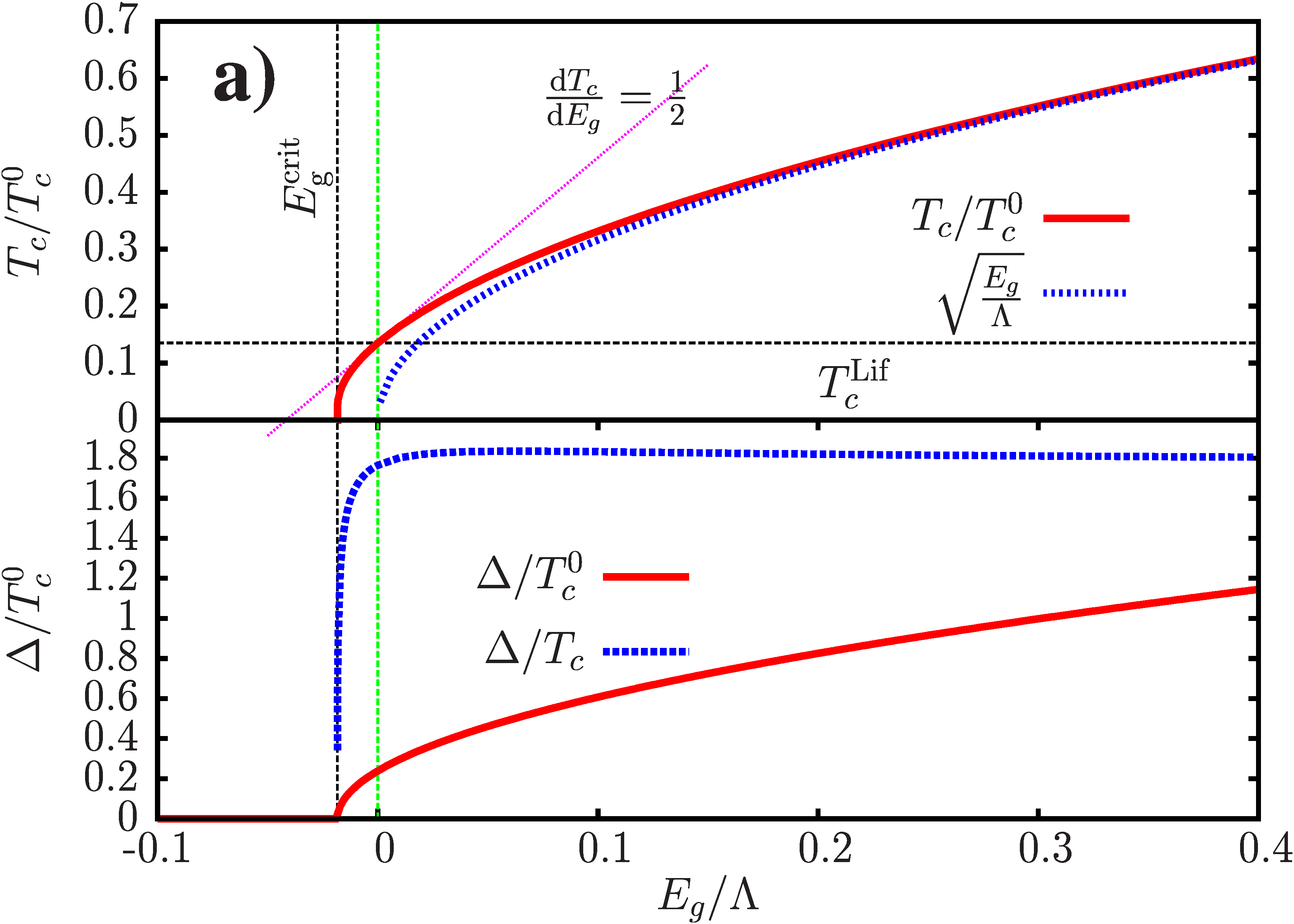}\\
\includegraphics[width=0.8\columnwidth]{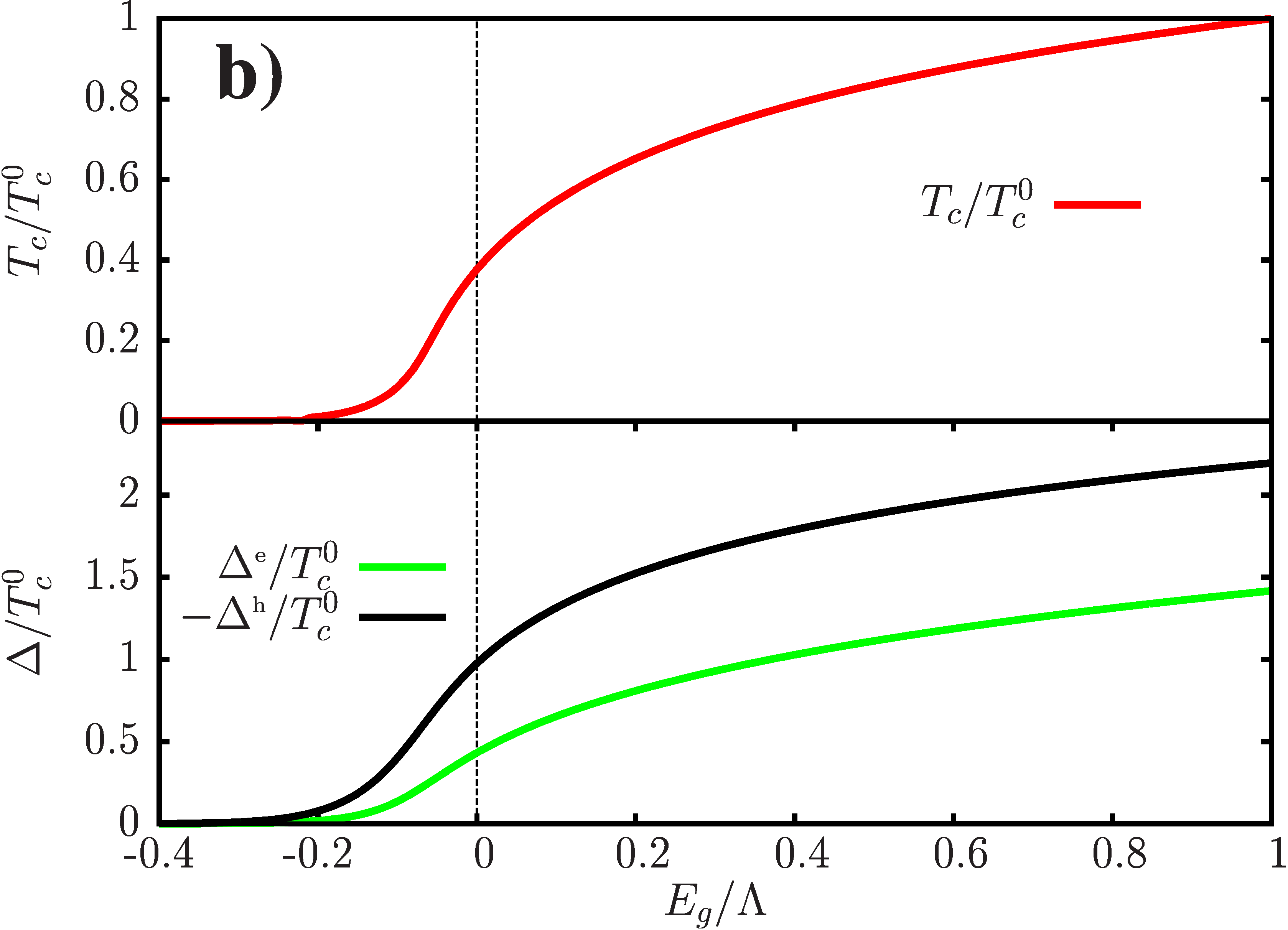}
\end{array}$
\caption{(Color online) (a)  $T_c$ and  gap as a function of $E_g$ for case(I)A. All the special lines are explained in the text. (b)  The same for case(I)B,   also discussed in Ref. \onlinecite{Bang}.  The normalization is with respect to $T_c^0$,  the critical temperature when $E_g>\Lambda$.  The dimensionless interactions were taken to be $v_{ph}=-0.5$ in (a) and
$v_{sf}=0.3$ in (b).}
   \label{fig:2}
\end{figure}

Accounting for the cutoff of available hole states at $E_g$ implies that  the  gap equation loses particle-hole symmetry and takes the form
\bea\label{eq:3}
1&=&-\frac{mV_{\text{ph}}}{2\pi}\left[\int^0_{-\Lambda}\frac{d\e}{2E}\tanh\frac{E}{2T} + \int^{E_g}_{0}\frac{d\e}{2E}\tanh\frac{E}{2T}\right],\nonumber\\
&&~~~~~~~~~~~~~~~~~~~~~~~~~~~~~~~~~~~~~~~~~~~~\text{for instance 2},\nonumber\\
1&=&-\frac{mV_{\text{ph}}}{2\pi}\left[\int^{-|E_g|}_{-\Lambda}\frac{d\e}{2E}\tanh\frac{E}{2T}\right]~\text{for instance 3},
\eea
To solve for $T_c$, we note that $E\rightarrow |\e|$. The solution of $T_c$ with $E_g$ is shown in Fig. \ref{fig:2}. One can get analytical expressions for some interesting  regimes:\\
The \emph{shallow band} ($E_g\lesssim\Lambda$) region gives  ($T_c\ll E_g$)
\beq\label{eq:shallow}
\frac{T_c}{T_c^0}=\sqrt{\frac{E_g}{\Lambda}},
\eeq
where $T_c^0$ is the  weak coupling critical temperature for $E_g>\Lambda$ and is given by $T_c^0 = \frac{2e^{\gamma}}{\pi}\Lambda e^{1/v_{\text{ph}}}$ ($\gamma$ is the Euler's constant) and $v_{\text{ph}} = mV_{\text{ph}}/2\pi$. We refrain from using the term `density of states' for $m/2\pi$ as it is usually reserved for states at the Fermi level.
 This has a physical relevance because close to the Lifshitz transition, the mass can be treated as constant within the pairing cutoff for a general dispersion.\\
Near the \emph{Lifshitz transition}($E_g\sim 0$) we get
\beq\label{eq:lift}
T_c=T^{\text{Lif}}_c+\frac{E_g}{2},
\eeq
where
\beq\label{eq:lift2}
T^{\text{Lif}}_c=T_c^0e^{\frac{1}{v_{\text{ph}}}}
\eeq
is the critical temperature at the Lifshitz point which is obtained by setting $E_g=0$ in Eq. \ref{eq:3}. The behavior of $T_c$ as a function of $E_g$ can be seen in the top panel of  Fig. \ref{fig:2}. It is clear that superconductivity in the system is suppressed  rapidly as $E_g$ falls below zero, as expected.
The gap on the incipient band is related to $T_c$ by the standard BCS ratio $\sim1.76$ until almost the incipient point $E_g=0$, before it drops drastically and vanishes at the critical $E_g^{\text{crit}}$.

$E_g^{\text{crit}}$ is the final important feature of the incipient solution for a single band where SC disappears before the lower cutoff $-\Lambda$ is reached. This is found by setting $T\rightarrow 0$ in Eq. \ref{eq:3} (instance-3). This immediately yields,
\beq\label{eq:Ec}
E_g^{\text{crit}}= -\Lambda e^{\frac{2}{v_{ph}}}.
\eeq
What this also implies is that, for a given $E_g<0$,  $v_{ph}$ cannot be made arbitrarily small and still obtain $T_c>0$, unlike the conventional BCS paradigm. This is the only case within weak-coupling where a threshold problem is encountered for the pairing interaction.
\subsection{Case (I)B}
The multiband scenario case(I)B with interband interaction $V_{sf}$ (the dimensionless interaction is $v_{sf}=\sqrt{N_em/(2\pi)}V_{sf}$ where $N_e$ is the Fermi level DOS of the electron band) also exhibits a similar strong suppression of $T_c$ (see  Fig. \ref{fig:2}(b) and Ref. \onlinecite{Bang}), but does not have a threshold. To see this, note that the gap equations for the two band problem are
\bea\label{eq:we}
&&\D_h=-V_{sf}L_e\D_e,~\D_e=-V_{sf}L_h\D_h;\nonumber\\
&&L_e=N_e\int_{-\Lambda}^{\Lambda} \frac{d\e}{2E_e}\tanh\frac{E_e}{2T},\nonumber\\
&&L_h=\frac{m}{2\pi}\int_{-\Lambda}^{E_g} \frac{d\e}{2E_e}\tanh\frac{E_e}{2T},
\eea
and the $T_c$ equation then reads
\beq\label{eq:intro}
1=V_{sf}^2L_eL_h.
\eeq
For the deep  incipient case, if $|E_g|\gg T_c$, only $L_e$ contains $\ln T_c$ and $L_h\sim \ln\Lambda/|E_g|$. This explains (1) why the effective pairing interaction now varies as $V^2\ln\Lambda/|E_g|$ as mentioned in the introduction; and (2) why $T_c$ (although strongly suppressed) exists for arbitrarily small $V$. The evolution of the two gaps with $E_g$ are also plotted in Fig. \ref{fig:2}.

\section{SC in incipient band - Case(II)}\label{Sec:case2}
We now switch to the discussion which presents the main message of this article: contrary to the prevalent  belief\cite{HongDingLiFeAs2015},  in the presence of well-stabilized SC, an incipient band can significantly enhance $T_c$.  In addition,  the induced SC gap on the incipient band  can be large.  We illustrate this by considering two cases which are motivated by some FeSC materials and will be discussed in detail in Sec. \ref{Sec:Disc}. These two cases differ essentially in the mechanism driving the   SC in the system  that exists in the absence of the incipient band. We start with case(II)A which is the generic case for FeSCs undergoing a Lifshitz transition.
\subsection{Case(II)A}
Our model here consists of one regular hole band at the $\Gamma$-point with Fermi level DOS $N_{h_1}$; one regular electron band forming two pockets at the $M$ points with Fermi level DOS $N_e$; and an incipient hole band ($h_2$) as modelled in Case(I). The interband pairing interaction  with a cutoff of $\Lambda$ drives SC in these bands. The origin of the pairing interaction is the  assumed presence of strong spin fluctuations  at $\textbf{Q}$, resulting from  particle-hole scattering between these bands as well as the incipient band. It is then reasonable to assume that the same interaction that stabilizes SC in the regular bands couples the incipient band to the rest of the system (namely the electron pockets). This consideration leads to the same  magnitude of the interband interaction between the electron band and the two hole bands (regular and incipient).   It will be useful to maintain generality  and distinguish the two interband interactions $V_{sf_1}$ and $V_{sf_2}$ connecting the electron band to the bands $h_1$ and $h_2$ respectively. Then the gap equations are:
\bea\label{eq:gaps}
\D_{{ e}} & =&-V_{sf_1}\D_{{ h}_{1}}L_{{ h}_{1}}-V_{sf_2}\D_{{ h}_{2}}L_{{ h}_{2}},\\
\D_{{ h}_{1}} & =&-2V_{sf_1}\D_{{ e}}L_{{ e}},\\
\D_{{ h}_{2}} & =&-2V_{sf_2}\D_{{ e}}L_{{ e}}.
\eea
where
\bea\label{eq:def}
L_{h_1} & =&\int_{-\Lambda}^{\Lambda}{\rm d}\varepsilon N_{h_1}\frac{{\rm tanh}\frac{E_{h_1}}{2T}}{2E_{h_1}},\nonumber\\
L_{e} & =&\int_{-\Lambda}^{\Lambda}{\rm d}\varepsilon N_{e}\frac{{\rm tanh}\frac{E_e}{2T}}{2E_e},\nonumber\\
L_{h_2} & =&\int_{-\Lambda}^{E_{{\rm g}}}{\rm d}\varepsilon \frac{m}{2\pi}\frac{{\rm tanh}\frac{E_{h_2}}{2T}}{2E_{h_2}},
\eea
$m$ is the mass of the incipient band. These equations can be rewritten as:
\bea\label{eq:def2}
\frac{\D_{h_2}}{\D_{h_1}}&=&\frac{V_{sf_2}}{V_{sf_1}},\nonumber\\
\frac{\D_{h_1}}{\D_{e}}&=&-2V_{sf_1}L_e,\nonumber\\
1&=&2L_e\left[V_{sf_1}^2L_{h_1} + V_{sf_2}^2L_{h_2}\right].
\eea
The same relations hold at $T_c$ with $L_{h_1}/N_{h_1}=L_e/N_{e}=\ln\frac{2e^{\gamma}\Lambda}{\pi T_c}$ and $L_{h_2}=\frac{m}{2\pi}\int_{-\Lambda}^{E_g}\frac{d\e}{2\e}\tanh\frac{\e}{2T_c}$. These equations in (\ref{eq:def2}) carry all the `universal' information central to our results:
\begin{itemize}
\item{ The first equation suggests that the gap induced on the incipient band is related to the ratio of the interband interactions. Recalling that this is the same interaction that couples $h_1$ and $e$ bands, we expect $V_{sf_1}/V_{sf_2}\approx 1$,  despite the fact that $h_2$ is incipient.  Differences in the orbital character of the bands  can easily tilt this ratio in either direction, but accounting for this is beyond the scope of this calculation. Thus we see that the induced gap is generically comparable to, and can in fact be larger than, the pre-existing gap.  Note also that this last point implies that, within the model, the hole band SC gap will be large until it disappears discontinously when $E_g$ passes through $-\Lambda$. Of cause a realistic (as compared to BCS) interaction will smear out this behavior.}
\item{The third equation tells us about the effect of the incipient band on the preexisting gap. In the absence of the incipient band (simulated by setting $V_{sf_2} =0$) we have $1=2V_{sf_1}^2L_{h_1}L_e$. Adding the positive definite $V_{sf_2}$ term forces the combination $L_eL_{h_1}$ to drop. The second equation then suggests that \emph{both} the $T=0$ electron and hole gaps are increased due to the presence of the incipient band.}
\item{The same arguments can be used to justify that $T_c$ is increased in the presence of the incipient band.}
\item{This model does not have an interaction  threshold for pairing. $T_c$ always exists.}
\item{The final piece of information contained in these equations is that the effect of SC on the incipient band   in this case is essentially the same as the effect on the regular hole band. The effect of the incipient band itself on the regular bands depends on the mass of the incipient band, such that lighter bands barely effect the gaps and $T_c$.   It is worth noting that neither $T_c$ nor the gap on the incipient band itself is likely to exhibit any discontinuous behavior at the Lifshitz transition.}
\end{itemize}

\begin{figure}[htbp]
\includegraphics[width=0.8\columnwidth]{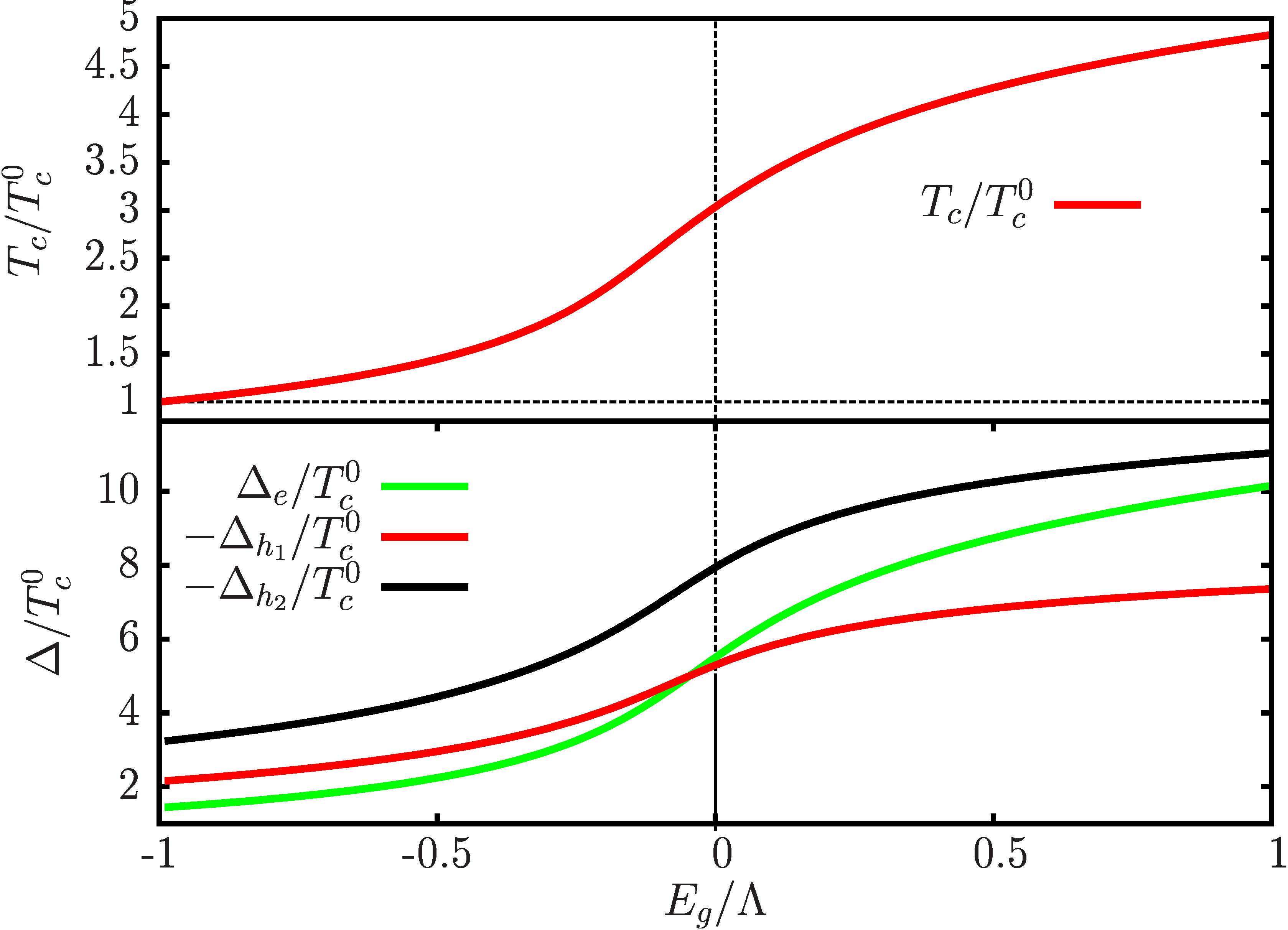}
\caption{(Color online)Case(II)A: 2D electron band with regular and incipient hole bands.   $T_c$ and gaps   as function of $E_g$. The normalization is with respect to $T_c^0$, the critical temperature when $ E_g<-\Lambda_{sf}$. Dimensionless interband interactions are $v_{sf_1}$=0.2 and $v_{sf_2}$=0.3.  Note that $\Delta_{h2}$ is the largest gap in the system.
}\label{fig:3}
\end{figure}

These equations can be solved for $T_c$ and for the $\D$'s at $T=0$ and  the solutions  are shown in Fig. \ref{fig:3}. As before, we can obtain analytical
results for special cases.    In the shallow band region ($0<E_g\lesssim \Lambda$),  if in addtion $E_g\gg T_c$,
\beq\label{eq:tc1}
T_c = T_c^0\exp\left[\frac{1}{\sqrt{2}v_{sf_1}}-\kappa -\sqrt{\kappa^2 + \frac{1}{2(v^2_{sf_1} + v^2_{sf_2})}}\right]
\eeq
where now we have defined  $T_c^0$ to be the transition temperature when $E_g<-\Lambda$; it is given by $T_c^0=\frac{2e^{\gamma}\Lambda}{\pi}e^{-1/\sqrt{2}v_{sf_1}}$,  with $v_{sf_1}=V_{sf_1}\sqrt{N_{h_1}N_e}$, $v_{sf_2}=V_{sf_2}\sqrt{mN_e/2\pi}$ and
\beq\label{wtf}
\kappa=\frac{\ln\frac{\Lambda}{|E_g|}}{4\left(1+ v^2_{sf_1} / v^2_{sf_2}\right)}.
\eeq
At the Lifshitz transition,
\beq\label{eq:tc2}
T^{\text{Lif}}_c = T_c^0\exp\left[\frac{1}{\sqrt{2}v_{sf_1}}\left(1-\frac{1}{\sqrt{1+ v^2_{sf_2}/ (2v^2_{sf_1}}}\right)\right].
\eeq
Near
 $T_c \ll |E_g| \lesssim \Lambda$
\beq\label{eq:tc3}
T_c = T_c^0\exp\left[\frac{1}{\sqrt{2}v_{sf_1}}+\kappa' - \sqrt{\kappa'^2 + \frac{1}{2v^2_{sf_1}}}\right],
\eeq
where
\beq\label{wtf2}
\kappa'=\frac{\ln\frac{\Lambda}{|E_g|}} {4v^2_{sf_1}/v^2_{sf_2}},
\eeq
and when the hole band $h_2$ becomes a vegetable ($E_g<-\Lambda$), we recover $T_c^0$.

\subsection{Case(II)B}
The toy model we choose here is the one where we have a regular electron band crossing the Fermi surface at the $M$ points with Fermi level DOS $N_e$. The SC is stabilized here via  an attractive electron-phonon mediated interaction. We then introduce an incipient hole band at the $\Gamma$ point (Fig. \ref{fig:1}). The pairing interaction  between the electron band and  this band can be thought of as being due to spin fluctuations and/or phonons.  For the moment, we nominally refer to the interactions between bands as originating from spin fluctuations. The microscopic origin of spin fluctuations in the presence of  just the incipient band is not obvious, but it is important to note that good Fermi surface nesting or even states at the Fermi surface are {\it not} required for a large static particle-hole susceptibility as appears in spin fluctuation pairing\cite{Kuroki}.  We  investigate the effect of these fluctuations on SC, coupling  the electron and hole band via $V_{sf}$. We further assume  that the cutoff scale  for spin fluctuations is larger than that for phonons,  i.e. $\Lambda_{sf}>\Lambda_{\text{ph}}$. Following  steps similar to those above, we first find the $T^0_c$ without the incipient band, given by $T_c^0=\frac{2e^{\gamma}\Lambda_{\text{ph}}}{\pi}e^{1/v_{\text{ph}}}$, where $v_{\text{ph}}=N_eV_{\text{ph}}<0$.

\begin{figure}[htbp]
\includegraphics[width=0.8\columnwidth]{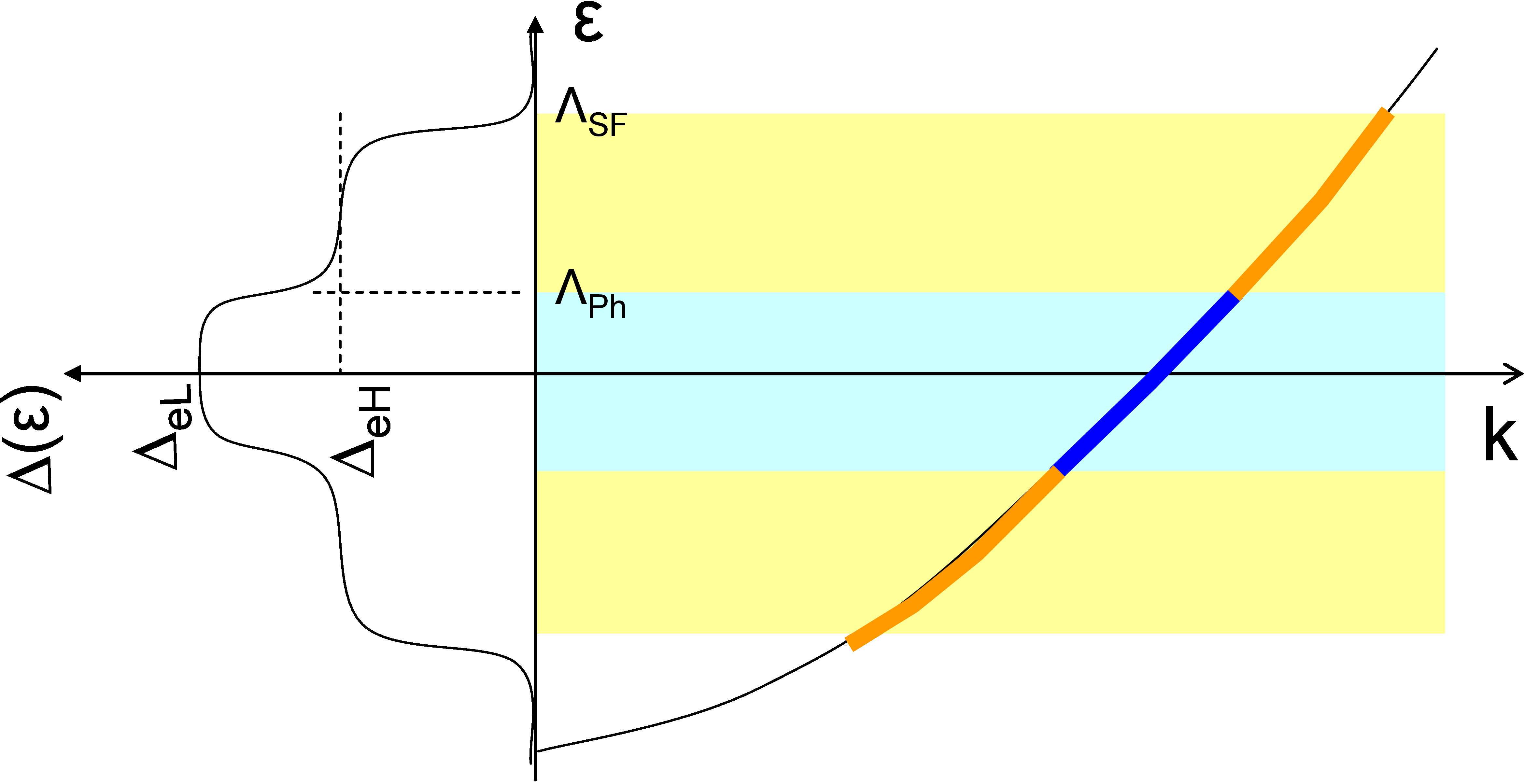}
\caption{(Color online) The gap structure on the electron band. The spin fluctuation interactions with the incipient hole band (not shown) have larger cutoff $\Lambda_{sf}$. The attractive phonon interactions within $\Lambda_{ph}$ causes the gap to be larger in the blue region.\label{fig:in}}
\end{figure}

\begin{figure}[htbp]
\includegraphics[width=0.9\columnwidth]{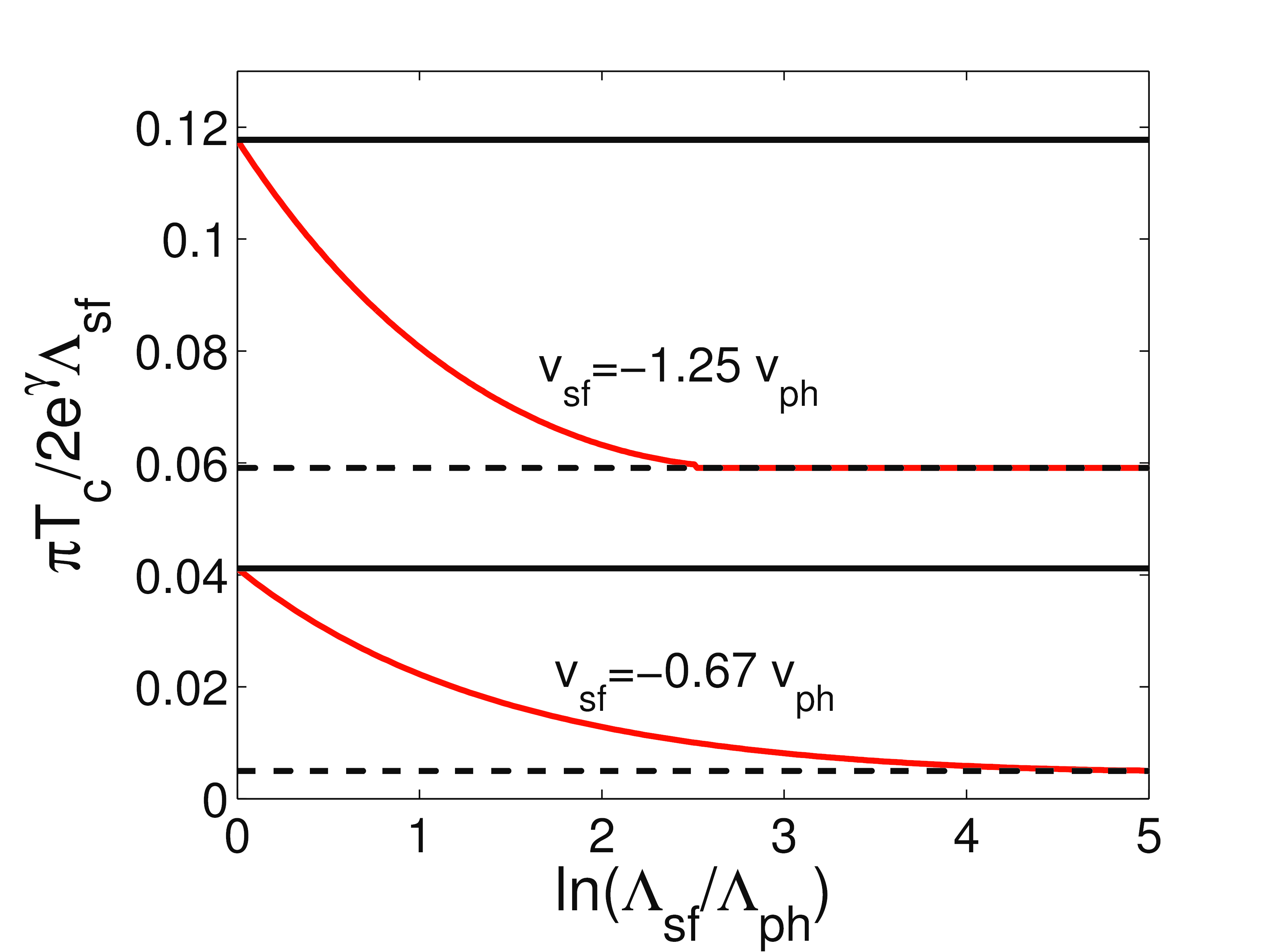}
\caption{(Color online)  Red solid lines: $T_c$ vs. $\ln \Lambda_{sf}/\Lambda_{ph}$ for two different values of $v_{sf}$, showing that $T_c$ increases as $\Lambda_{ph}$ is increased all the way up to $\Lambda_{ph}=\Lambda_{sf}$.
For each $v_{sf}$, dashed  black lines correspond to  $T_c$ for $\Lambda_{ph}=0$, while solid black lines correspond to
  $T_c$ for $\Lambda_{ph}=\Lambda_{sf}$.  Here $v_{ph}=-0.2$.}\label{fig:inter}
\end{figure}

Some care is needed in formulating this problem due to the presence of different cutoff scales for the pairing interactions. The main point of departure from  conventional BCS modelling is that the gap on the electron pocket is expected to vary at energy scales of $\Lambda_{\text{ph}}$. In the spirit of the Anderson-Morel model\cite{Anderson1960,Anderson1961}, we account for this effect by letting the otherwise constant electron gap to acquire different values $\D_{eL}$ for $|\e|<\Lambda_{ph}$ and $\D_{eH}$ for $\Lambda_{ph}<|\e|<\Lambda_{sf}$ ($L,~ H$ stand from low and high energy respectively, see Fig. \ref{fig:in}).
 Note that allowing $\Delta$ to vary with energy is outside the BCS approximation and high energy renormalizations may be relevant for a quantitative estimate, which is outside the scope of this work.

$V_{ph}$ is only felt by the electron band up to $\Lambda_{ph}$. Incorporating these into the gap equations,
we arrive at the following:
\bea\label{eq:gapph}
\D_{eL}&=&-V_{ph}\D_{eL}L_{eL}-V_{sf}\D_hL_h,\nonumber\\
\D_{eH}&=&-V_{sf}\D_hL_h,\nonumber\\
\D_h&=&-2V_{sf}(\D_{eL}L_{eL} + \D_{eH}L_{eH}),
\eea
where
\bea\label{eq:def3}
L_{eL}&=&2\int_{0}^{\Lambda_{ph}} d\e N_e \frac{\tanh\frac{E_{eL}}{2T}}{2E_{eL}},\nonumber\\
L_{eH}&=&2\int_{\Lambda_{ph}}^{\Lambda_{sf}} d\e N_e \frac{\tanh\frac{E_{eH}}{2T}}{2E_{eH}},\nonumber\\
L_{h}&=&\int_{-\Lambda_{sf}}^{E_g} d\e \frac{m}{2\pi} \frac{\tanh\frac{E_{h}}{2T}}{2E_{h}}.
\eea
We may rewrite these equations as:
\bea\label{eq:def4}
&&\D_{eH}=-V_{sf}\D_hL_h,\nonumber\\
&&(1+V_{ph}L_{eL})\D_{eL}=-V_{sf}\D_hL_h,\nonumber\\
&&(1+V_{ph}L_{eL})\left(\frac{1}{2V_{sf}^2}-L_hL_{eH}\right)=L_{eL}L_h.
\eea
We immediately see that, quite generally, from the first equation $\D_{eH}\D_h<0$; from the third equation, if $V_{sf}$ is introduced perturbatively, then $1+V_{ph}L_{eL}>0$ requiring $\D_{eL}\D_h<0$. Note that the introduction of $V_{sf}$, requires $1+V_{ph}L_{eL}=0\rightarrow1+V_{ph}L_{eL}>0$. Then, $V_{ph}<0$ suggests that $T_c$ and the $T=0$ gap must increase. Thus we see   that the introduction of the repulsive spin fluctuation
coupling to the hole band, normally assumed to be a completing    mechanism\cite{JishiScalapino}, actually aids the electron-phonon SC  in this case. This is the core of the bootstrapping effect described in the introduction.

In order to understand the effect of relative ratio of the cutoffs for the two mechanisms, let us focus, for simplicity, on the regular band case where $E_g>\Lambda_{sf}$. We define $v_{sf}\equiv\sqrt{N_hN_e}V_{sf}>0$ and $v_{ph}\equiv N_eV_{ph}<0$. The solution to $T_c$ for any $\Lambda_{sf}/\Lambda_{ph}$ is given by
\begin{widetext}
\beq\label{eq:wide}
\ln\frac{2e^{\gamma}\Lambda_{sf}}{\pi T_c}=\frac{1}{\sqrt{2}v_{sf}}\left[\frac{-(r+l(2-rl)) + \sqrt{(r+l(2-rl))^2+4(1-l^2)(1-rl)}}{2(1-rl)}\right]+ \frac{l}{\sqrt{2}v_{sf}},
\eeq
\end{widetext}
where $l\equiv \sqrt{2}v_{sf}\ln\frac{\Lambda_{sf}}{\Lambda_{ph}}$, $r=-\frac{v_{ph}}{\sqrt{2}v_{sf}}$. We now show that this correctly reduces to the well known cases when $\Lambda_{sf}\rightarrow\Lambda_{ph}=\Lambda$ and when $\Lambda_{ph}\rightarrow 0$. It is clear from Eqs. \ref{eq:gapph} and \ref{eq:def3} that  in the limit $\Lambda_{sf}\rightarrow\Lambda_{ph}=\Lambda$, so that $L_{eH}\rightarrow 0$,  there is no `phase space' for $\Delta_{eH}$.  This then  correctly reduces to the usual model with 2-band whose $T_c$ is given by
\beq\label{eq:lim1}
\ln\frac{2e^{\gamma}\Lambda_{ph}}{T^{\text{2Band}}_c}=\frac{1}{\sqrt{2}v_{sf}}\left[\frac{-r + \sqrt{r^2 + 4}}{2}\right].
\eeq
The same is achieved by setting $l=0$ in Eq.\ref{eq:wide}. In the other limit, $\Lambda_{ph}\rightarrow 0$, we note that $L_{eL}\rightarrow 0$ and $L_{eH}\rightarrow \ln \frac{2e^{\gamma}\Lambda}{\pi T_c}$ (or equivalently $l\rightarrow \sqrt{2}v_{sf}\ln \frac{2e^{\gamma}\Lambda}{\pi T_c}$). This means that Eq. \ref{eq:wide} needs to be solved for $\ln \frac{2e^{\gamma}\Lambda}{\pi T_c}$. In doing so, using $1-rl\neq0$ we end up  with  $l=1$ or $\ln \frac{2e^{\gamma}\Lambda}{\pi T_c}=1/\sqrt{2}v_{sf}$. This is the well known $T_c$ for the 2 band toy $s\pm$ SC model.

Having convinced ourselves that the model reproduces the two limits of applicability, we now look at the general solution, plotted in  Fig. \ref{fig:inter}. As expected, $T_c$  generally increases when $v_{sf}$ is increased. There is however a possibly interesting interplay with the ratio $\Lambda_{sf}/\Lambda_{ph}$: as we increase $\Lambda_{ph}$, the $T_c$ increases (all the way up to  where the two cutoff's are comparable). It suggests that the presence of both mechanisms should help increase $T_c$.

\begin{figure}[htbp]
\includegraphics[width=0.8\columnwidth]{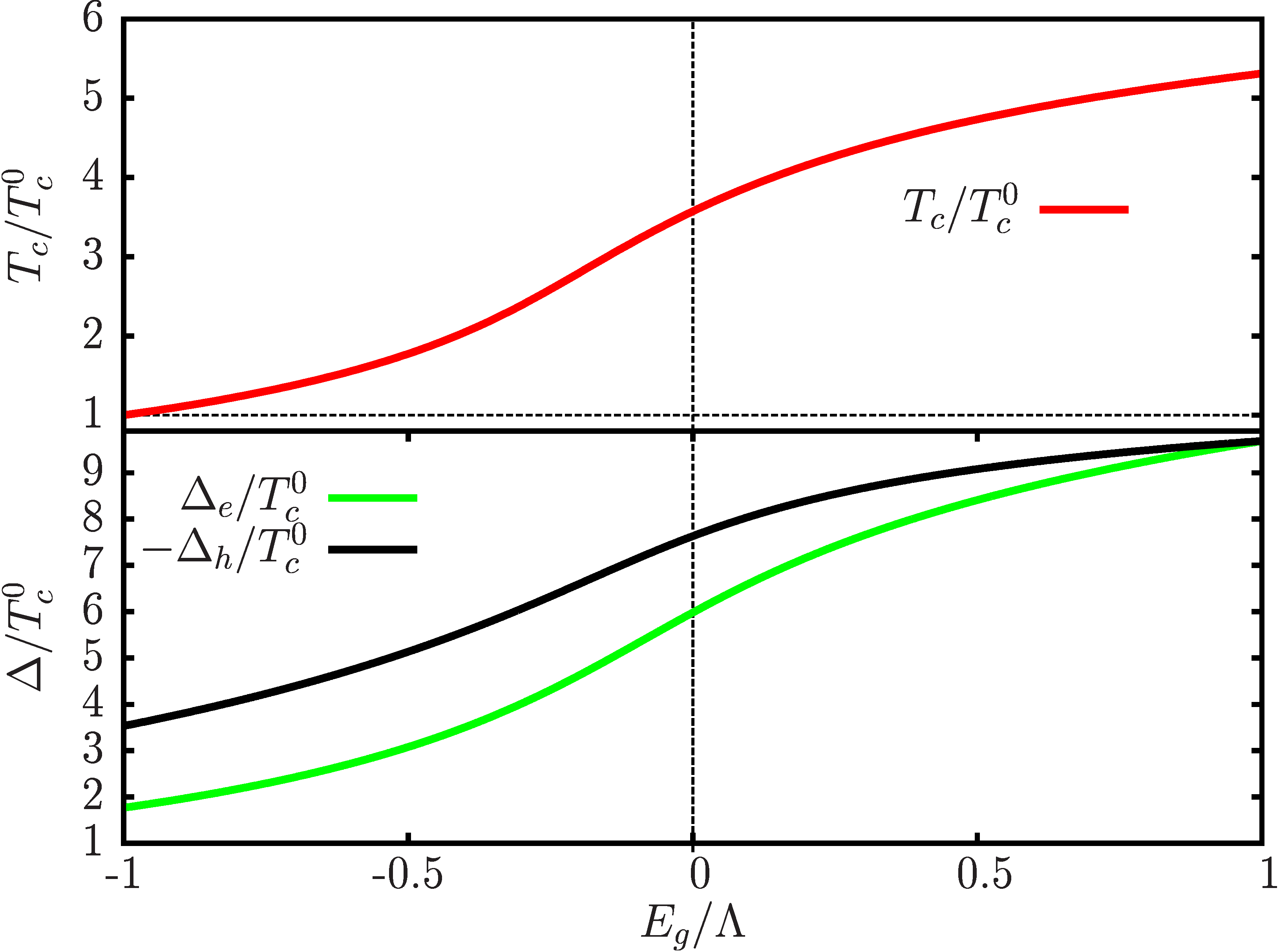}
\caption{(Color online)  Case(II)B: 2D electron and hole band.  $T_c$ normalized to $T_c^0$,  the transition temperature when $E_g<-\Lambda_{sf}$,  and gaps as a function of $E_g$, normalized to $\Lambda_{sf}=\Lambda_{ph}$. Dimensionless interactions are  $v_{ph}$=-0.3,
and $v_{sf}$=0.3.
}\label{fig:4}
\end{figure}

Returning to the incipient problem, we wish to study $T_c$ and the gaps on the two bands as a function of $E_g$. We perform the usual change with $N_h\rightarrow m/2\pi$ and work in the limit $\Lambda_{sf}\rightarrow \Lambda_{ph}$. These results are plotted in Fig. \ref{fig:4}. We see that not only is  the $T_c$ is enhanced due to the presence of the incipient band as expected from the above discussion,  but the crossover through the Lifshitz transition is considerably less abrupt than in Cases (I).  We refer to this key result in our discussion of FeSe monolayers on STO, see Sec. \ref{Sec:Disc}.

\subsection{Effect of 3 dimensionality of the incipient band}

\begin{figure}[htbp]
\includegraphics[width=0.8\columnwidth]{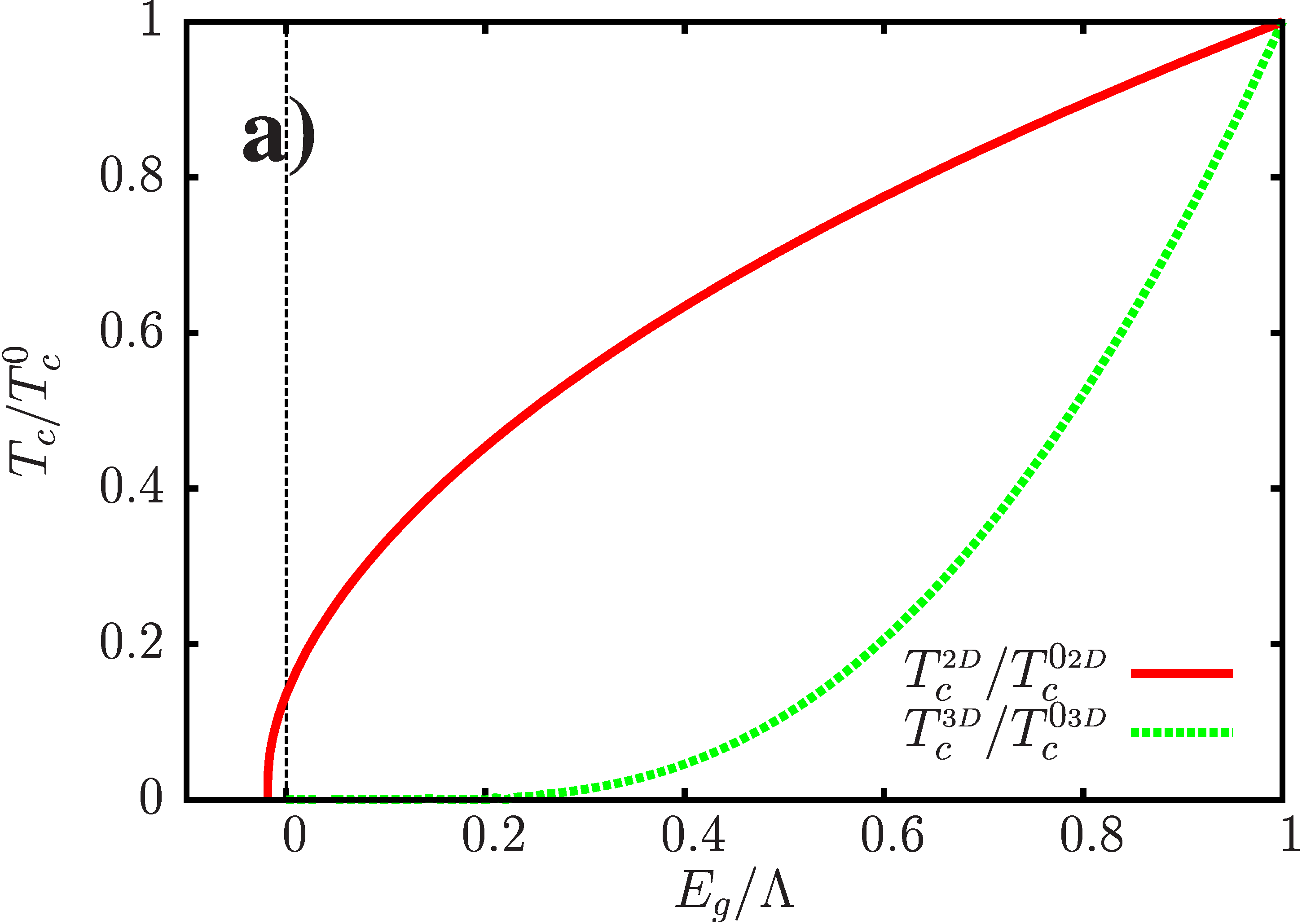}
\includegraphics[width=0.8\columnwidth]{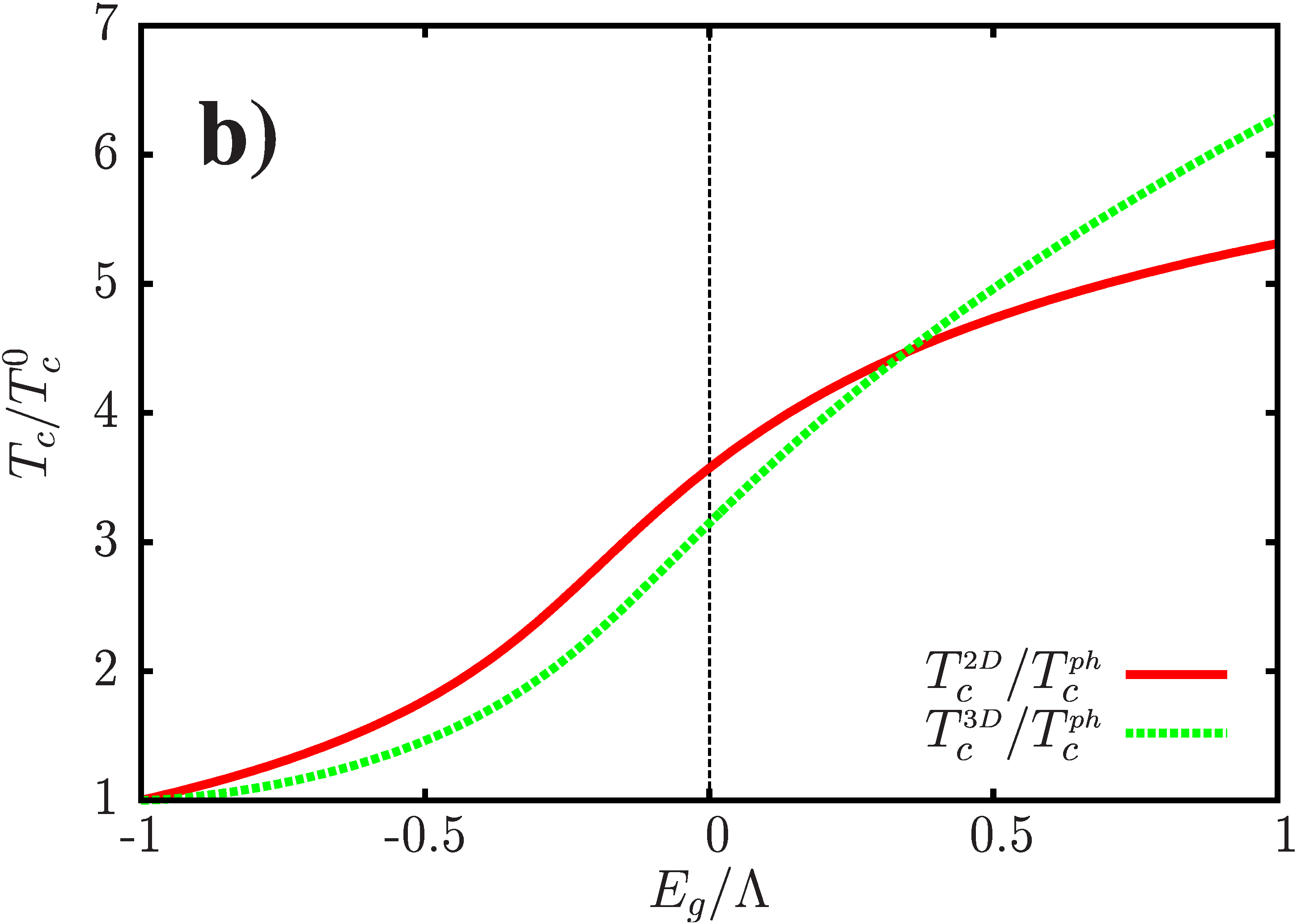}
\caption{(Color online) Comparison of 2D (red)  and 3D (green) scenarios.
(a) Case(I)A  $T_c/T_{c}^0$ vs $E_g/\Lambda$, where $T_c^0$ is the value of $T_c$ at $E_g=\Lambda$. The dimensionless phonon interaction was taken as $v_{ph}=-0.5$ for 2D and $v_{ph}=-0.2\sqrt{2}$ for 3D, with DOS ratio
$m/(2\pi N_h^{\scriptscriptstyle \rm 3D})=1.$
 (b) Same, but for Case (II)B, with
 $v_{ph}=-0.3$ and $v_{sf}=0.3$ for both 2D and 3D.   For this plot, the special case $\Lambda_{ph}=\Lambda_{sf}=\Lambda$ was adopted .
\label{fig:55}}
\end{figure}

We have so far only addressed 2D systems where the conversion of the phase-space $\bk$ integral to energy integral,
near the Lifshitz point, was done via $\int\frac{d^2k}{(2\pi)^2}=\frac{m}{2\pi}\int d\e$ for all energies
(the  constant density of states for parabolic bands).
This property changes in 3D since for a hole band with dispersion $-k^2/2m + E_g$,
\bea
N_h(\e)=N_h^{\rm{\scriptscriptstyle{3D}}}\text{Re}\sqrt{2\frac{E_g-\e}{\Lambda}}\,,
\eea
where $N^{\rm \scriptscriptstyle 3D}_h$ is given by  $a\sqrt{\frac{\Lambda}{2}}$, where $a\equiv\frac{(2m)^{3/2}}{4\pi^2}$.
We give details in this less transparent case in the Appendix.
Below, we give a qualitative discussion with the focus on the question whether the
previous results for a 2D hole band are substantially modified.

The weighting factor $~\text{Re}\sqrt{E_g-\e}$ near the top of the band in the energy space proves harmful for the $T_c$ in the one band incipient
case(I)A,  as can be seen from Fig.~\ref{fig:55} (a) where we compare the 2D (red curve) with the 3D version (green dashed curve).
 It is clear that $T_c$ in 3D is suppressed significantly due to the depletion of the DOS relative to the 2D case within weak coupling as the band becomes incipient.   In fact, there is no SC in the 3D incipient band case for case (I)A.
SC is present for a shallow band for any strength of attractive interaction
in the form of the BCS essential singularity  $T_c\sim\exp(-1/a\sqrt{E_g}V_{ph})$, but completely suppresed for
an incipient band.
This result originates in the 3D analog of the integral of Eq.~\eqref{eq:3}. The additional
square root
that removes the singular nature of the integrated kernel
as $T_c\rightarrow 0$ as compared to the case in 2D, where the kernel is $~\rm{tanh}(\e/2T_c)/\e$.
Thereby, the influence of $T_c$ on the value of the integral is reduced in 3D and no weak coupling solution is possible at or
beyond the Lifshitz transition. Only if we allow for strong coupling SC in the sense that $T_c$ is larger
than the cutoff $\Lambda$ do we find SC for a 3D incipient band.

Thus, the question arises if a similar conclusion holds in the multi-band scenarios discussed in this work,
 i.e. whether or not SC is strongly suppressed by such 3D effects.
Even without an explicit calculation, we see that the log singularity of the BCS integral is  again lifted.
The value of the integral can be large while the influence of $T_c$ on this value is small.
 In order to compare 2D and 3D, we choose a reference point such that the 3D DOS equals the 2D DOS at $E_g=\Lambda/2$. We now calculate the $T_c$ enhancement of
a phonon mediated SC with a 3D incipient hole band and compare the result with case(II)B in Fig.~\ref{fig:55}(b).
We observe a rather moderate reduction of the enhancement in the incipient region with a 3D hole band (dashed green curve) as compared to the 2D case (red curve).
In the Appendix, we show further results where we repeat the calculations of the main text with a 3D hole band.
Similar to Fig.~\ref{fig:55}, we observe that a 3D hole band can bootstrap SC at the Fermi level almost as effectively as a 2D band.

\section{Discussion}\label{Sec:Disc}
The main result of our analysis is that, for the Fe-based superconductors,  the appearance of superconductivity on an incipient band is a rather natural consequence of multiband pairing, rather than  an indication of strong coupling physics.   Here we discuss how our results relate to various controversies in the field for particular materials at the present time, in a rather simplified way that neglects various complications, such as the exact number of bands, orbital character, etc.    In each of these cases, more detailed theoretical work is needed to address the issue of the consequences for pairing of incipient bands in the system, since the vast majority of the detailed calculations have assume pairing only at the Fermi surface.

{\it LiFeAs.} The fascinating experiment which revitalized this discussion, Miao et al. \cite{HongDingLiFeAs2015}, showed the  persistence of large gap on a hole band as it underwent a Lifshitz transition upon Co doping.    The lack of any significant signature of this Lifshitz transition in either $T_c$ or the ARPES gap magnitude suggested to the authors of this work that
weak coupling physics, which relies on Fermi surface interactions, could not be at play.  They furthermore argued that induced superconductivity, due to the interactions between the bands at the Fermi surface and ``proximity coupled" in momentum space to the incipient band, could not be occurring because  the gap observed there was the largest in the system. Subsequently, Hu et al.\cite{FCZhang2015}  considered a multiband situation superficially similar to our case (II)B, and found that gaps the size observed in the experiment required strong coupling, i.e. dimensionless interband interactions of order 1, and in addition reported that their equations required a critical interaction strength to generate a finite $T_c$.   They claimed that their results qualitatively supported the conclusions of Ref. \onlinecite{HongDingLiFeAs2015}.

\begin{figure}[htbp]
\includegraphics[width=0.8\columnwidth]{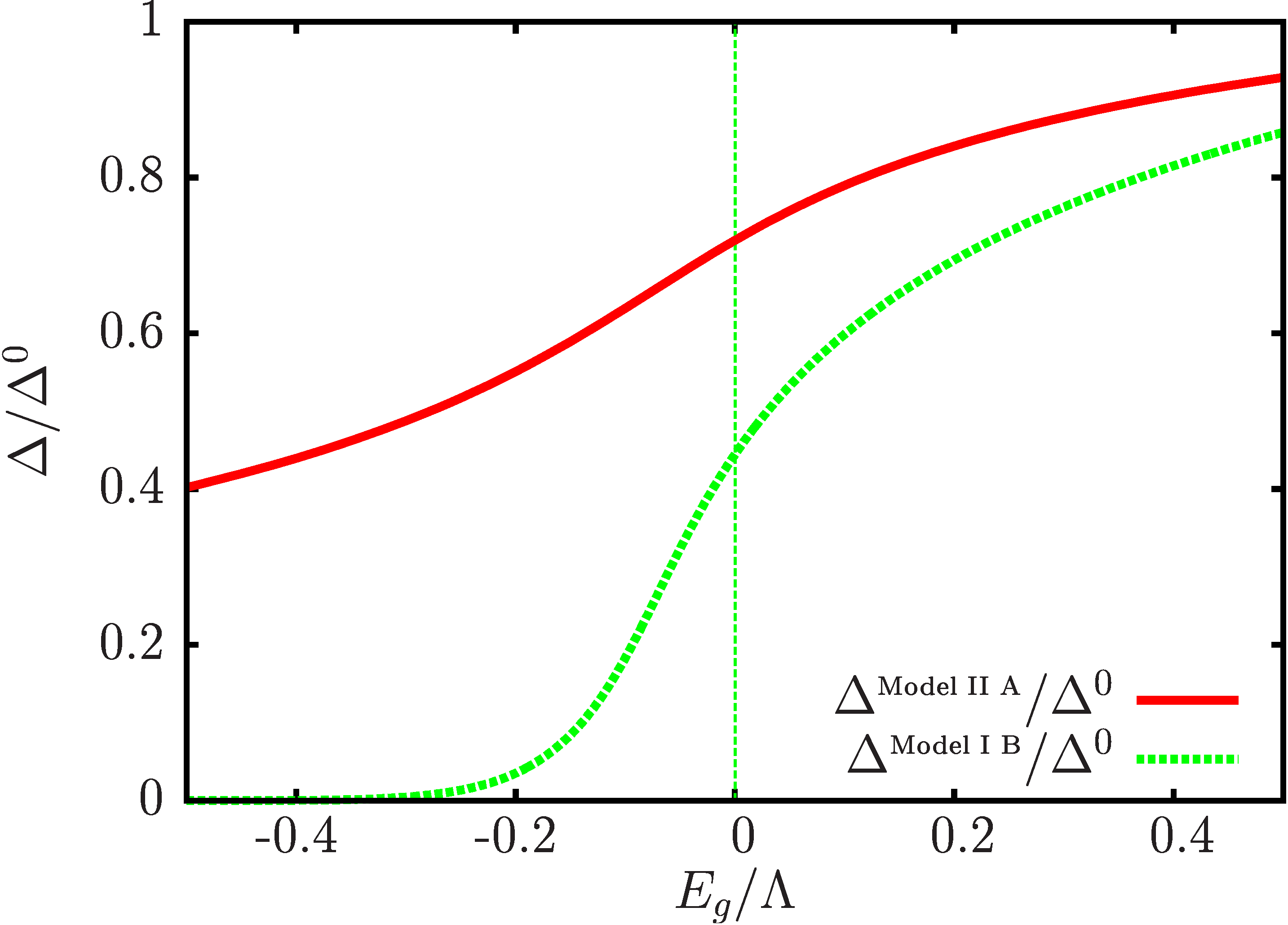}
\caption{(Color online) Comparison of the incipient hole gap in case(I) (green) and case(II) (red) taken from Fig.~\ref{fig:2} (b) and \ref{fig:3}, respectively.  Gaps are normalized to their value $\Delta_0<0$ at $E_g=\Lambda$.}
\label{fig:5}
\end{figure}

On the other hand, we have demonstrated that  effects of the type observed by Miao et al.\cite{HongDingLiFeAs2015} are rather easy to generate in a case (II)A situation.  This is certainly characteristic of LiFeAs, which has electron and at least one, possibly two hole pockets at the Fermi level\cite{Borisenko_Symmetry12,Umezawa12,Chi2014} even with substantial Co doping.
In Fig.~\ref{fig:5}, we compare the gap on the incipient band as a function of $E_g$ for  cases (I)B and (II)A. While we have already seen in Fig.~\ref{fig:3} that, depending on the ratio of the interactions and the DOS, the gap on the incipient band can be the largest in the system, we now clearly see that it is only weakly suppressed as the band sinks below the Fermi level.
We furthermore disagree with the conclusions of Hu et al.\cite{FCZhang2015}, because it appears to be based on an incorrect formulation of the  multiband pairing
problem. In Eq. (6)  of their article,
they include the interband interaction $V_2$ into the intraband kernel. In doing so,
the problem they solve actually maps to our intra-band pairing scenario (case(I)A) and
hence they see the threshold for the pairing interaction.  One can easily check that, as a result of this, their
gap equation does not reduce to the classic two-band $s\pm$ superconductivity as  discussed, e.g. in
Ref. \onlinecite{mazin08} when the intraband interaction $V_1=0$, whether or not one of the bands is
incipient.

The  rough conclusions that we present here may be of considerable relevance for theoretical calculations of the pairing state of LiFeAs.   Because it is nonmagnetic, with no obvious nesting, and because high-quality ARPES data (including precise measurements of anisotropic gaps on various Fermi surface sheets) have been available due to the nonpolar surfaces of this material,  LiFeAs has been perceived as something of a challenge by theorists.  Several proposals have been made, both based on DFT-derived Fermi surfaces\cite{ThomaleLiFeAs}, or on the rather different ARPES-determined Fermi surfaces\cite{Wang13,Ahnetal,Yin2014,KontaniLiFeAs}.  At issue has been the size of the gaps on the rather small inner hole $xz/yz$ Fermi surface pockets reported by ARPES which are those which undergo Lifshitz transitions upon electron doping.   Empirically, these gaps are the largest in the system, and this property is retained upon electron doping, even when the bands responsible fall below the Fermi level.    The calculations in question all considered pairing only at the Fermi level, and generally agreed on the gap functions obtained for the electron and outer hole pockets, but disagreed on the sizes of the gaps on the inner hole bands.    In some cases, good agreement with the gaps on the smaller hole pockets were found, based on claims of improved calculational schemes\cite{Yin2014,KontaniLiFeAs}.  In the case of the only fully 3D spin fluctuation pairing calculation, Ref. \onlinecite{Wang13},  the gaps on these small pockets were found to be too small compared to experiment, and the authors speculated that this might be due to the neglect of states away from the Fermi level, including states in incipient bands.  Our calculations here suggest that such effects could indeed be important, and it may be that for such systems Eliashberg or other calculational schemes which account for the dynamics of the pairing interaction are required.

{\it FeSe monoloayers on STO.}
While the lattice parameters of the FeSe monolayers grown on STO, with $T_c$'s of 70K or higher, are a few percent larger than that of the bulk,  it has proven difficult to reproduce the experimental Fermi surface by DFT calculations for a 2D FeSe layer,  accounting only for the strain.
Most researchers believe that the O vacancies in the STO play an important role by  electron-doping the
FeSe monolayer and thereby pushing down the $\Gamma$-centered hole band\cite{Gong2014} .  Another clue to the physics of these systems, and the influence of the substrate, was recently provided by ARPES measurements\cite{Shen_FeSe-STO}, which indicated via the observation of ``replica bands"  the presence of a strong electron-phonon interaction, probably originating from the substrate\cite{Shen_FeSe-STO}.  It has recently been argued that the electron-phonon interaction must be quite peaked near momentum transfer ${\bf q}$=0 to explain this observation\cite{SJohnston2015}, supporting the basic scenario for high-$T_c$ proposed in the Refs.~\onlinecite{Shen_FeSe-STO} and \onlinecite{Lee2015}.

Considering only the bands at the Fermi surface, the high-$T_c$ in this system and the form of the order parameter  are  puzzling.  We do not expect electron-phonon interactions in the FeSe  to be strong enough to explain a $T_c$ of 70K or above\cite{Li_phonon}, such that a simple $s$-wave from attractive interactions alone seems unlikely, even if boosted by soft STO phonons.   The forward scattering nature of the essential phonon processes then means that phonons cannot  contribute to the interband interaction.   On the other hand, the   spin fluctuation interaction by itself  should lead naively to nodeless $d$-wave (since $\chi({\bf q},\omega)$ will be peaked at the wave vector connecting the electron pockets), as in the arguments given for alkali-intercalates\cite{Fangdwave,Maierdwave}.   There are some indications that the system does not have a sign-changing order parameter, however.  For example, STM measurements by Fan et al. \cite{Fanetal_impurities_Fe-STO15} show a full gap which is suppressed only by magnetic impurities, similar to a ``plain" $s$-wave superconductor.  Note that these arguments, if correct, would also rule out states of the ``bonding-antibonding $s$-wave" type\cite{HKMReview2011}, which we do not discuss here.

\begin{figure}[htbp]
\includegraphics[width=1.0\columnwidth]{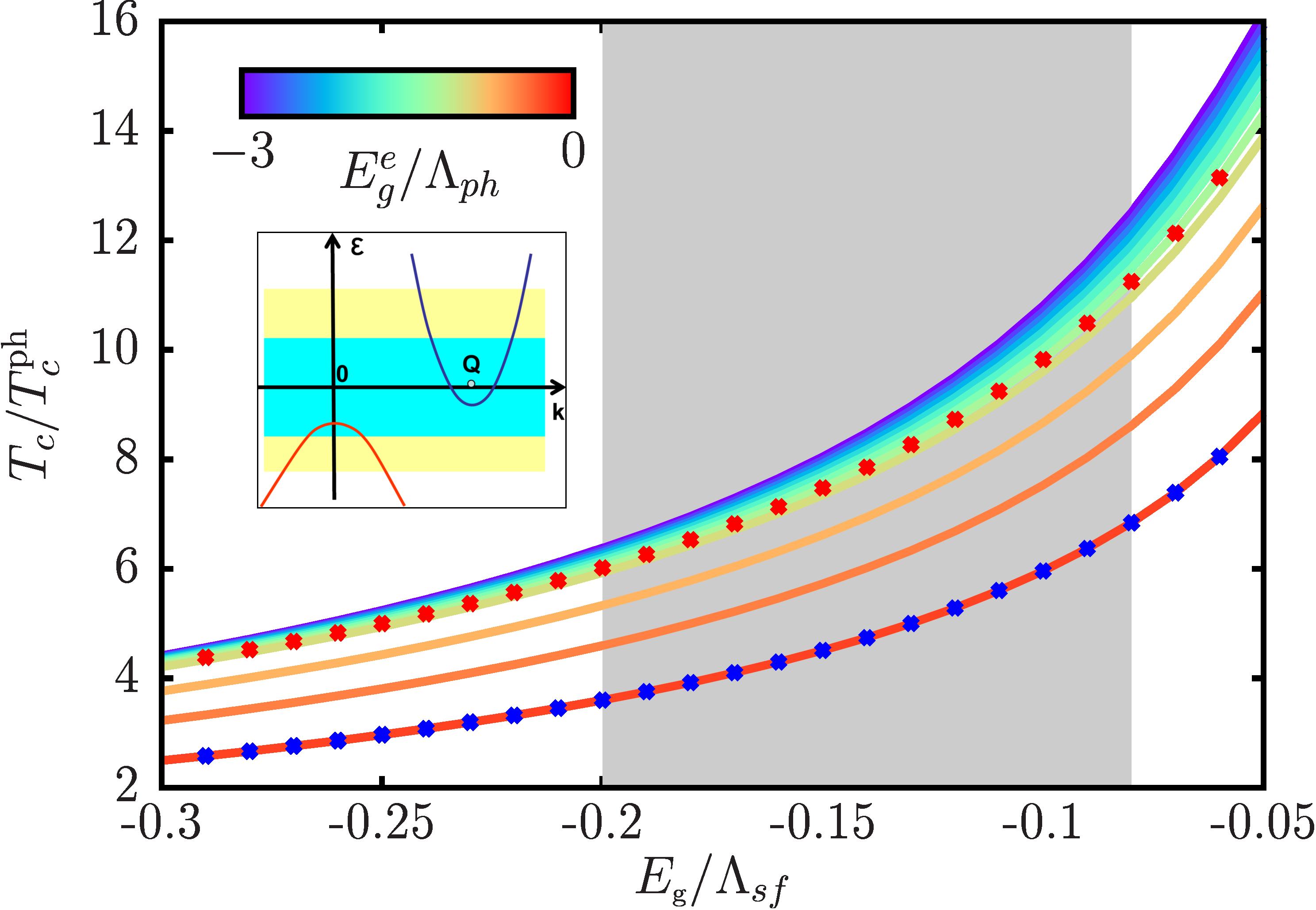}
\caption{(Color online)
 $T_c$ as a function of the band edge $E_g$ of the incipient hole band for
several band extrema for the electron band $E_{e}$.
Big red dots highlight the curve $T_c(E_g)$ where $E_{e}=-\Lambda_{ph}$ and big blue
dots $E_{e}=0$ where the electron band reaches the Fermi level.
The experimental situation of a shallow band (-$\Lambda_{ph}<E_{e}<0$) is in between
these curves.
 The shaded gray region is a range of $E_g/\Lambda_{sf}$ for $E_g=80$meV and the SF cutoff ( $\Lambda_{sf}=400$ to $1000$ meV) that is a rough estimate based on calculations for bulk FeSe \cite{EssenbergerSFFeSeArXiv}.
  We use $v_{ph}=-0.2,$ $v_{sf}=0.25$, $\Lambda_{sf}=600$ meV and $\Lambda_{ph}=100 $ meV.  For these
 parameters $T_c^{ph}=9K$.
\label{fig:FeSeModel}}
\end{figure}

The arguments in this paper favor the ``dark horse" candidate for pairing,
the incipient $s_\pm$ state, with a large gap  magnitude on both the electron pockets at the  Fermi surface band and the incipient hole band well below it.
The spin fluctuations have been shown capable of substantially enhancing a weak phonon $T_c$.
 In order to account for the experimental situation we slightly modify the model case(II)B.
It was shown that the hole band is pushed below the bottom of the electron band\cite{Shen_FeSe-STO},
but the presence of the replica band suggests that even the hole band is in the range of the phonon cutoff.
Thus, we consider a shallow electron band  (band minimum $E_g^e$) together with an incipient hole band, but otherwise similar, situation as in case(II)B.
 We include phonon coupling in the part of the incipient hole band within the phonon cutoff.
The model is shown in the inset of Fig.~\ref{fig:FeSeModel}.
To illustrate the possibility of incipient spin fluctuation bootstrap more concretely,
we plot in Fig. \ref{fig:FeSeModel} the possible $T_c$ enhancements over a  phonon bare critical
temperature $T_c^{ph}$  that one would obtain in a na\"{i}ve calculation, i.e. the $T_c$ in the absence of interband spin fluctuations and disregarding all band edge effects ($E_g^e=-\Lambda_{ph}$).
The cutoff in spin fluctuation pairing theory is ill-defined, but may be roughly identified with the energy scale
 of the spin fluctuation Eliashberg function in Ref.~\onlinecite{EssenbergerSFFeSeArXiv} for bulk FeSe.
This Eliashberg function has appreciable weight for energies as high as 800meV and a peak at 600meV.
We account for this with a rough estimate represented by the gray shaded area in Fig.~\ref{fig:FeSeModel} that highlights the range of $E_g/\Lambda_{sf}$
for $E_g=80$meV and $\Lambda_{sf}=400$ to $1000$meV.

  For the choice of parameters used in the figure, $T_c^{ph}$ is  9K, and that the gray region suggests that enhancements of order  6-12  with respect to $T_c^{ph}$  are possible.      Note however, that this range is quite sensitive to the choices of interactions
and the ratio of the cutoffs, which  are poorly known, so the reader should not take the numbers particularly seriously.  Our message is simply that a weak bare phonon interaction can be used to create a large $T_c$, even with a spin fluctuation interaction which may be weakened by the incipient band.

Of course the $s_\pm$  state found here naively has the same difficulty with the results of Ref. \onlinecite{Fanetal_impurities_Fe-STO15}.  However, since impurities scatter elastically, one expects substantial suppression of pairbreaking effects due to the gap sign change in incipient band pair systems.  This question is currently under active investigation.

{\it FeSe intercalates.}  Here we intentionally lump together, without particularly good justification, a) alkali-doped FeSe intercalates like KFe$_2$Se$_2$,  b) ammoniated FeSe intercalated like Li$_{0.56}$(NH$_2$)$_{0.53}$(NH$_3$)$_{1.19}$Fe$_2$Se$_2$\cite{AmmoniaRichJACS}; and c)  recently discovered lithium iron selenide hydroxides Li$_{1-x}$Fe$_{x}$(OH)Fe$_{1-y}$Se. a) and c) have been shown to have Fermi surfaces without $\Gamma$-centered hole pockets, similar to the FeSe/STO monolayers\cite{Zhao2015}.    There are no ARPES Fermi surface measurements of the materials in category b) to our knowledge, due to sample volatility, but it seems reasonable to assume since FeSe interlayer distances are comparable, and $T_c$'s similar (of order 40K for a),b) and c)), that they may be in this class.

Since $T_c$ is not as high as in the FeSe/STO monolayers, it is tempting to speculate that these systems must all belong to class (I) B.   That is, in these systems we have no evidence (to our knowledge) that the electron-phonon interaction plays any exceptional role; we assume, therefore, that it may be neglected, leaving a strongly
suppressed incipient $s_\pm$ superconducting channel to compete with what  should be a much more robust
$d$-wave interaction present in all systems\cite{Maiti2011}.     Ultimately all case (I)B systems   should be $d$-wave as well.  In some systems, evidence against $d$-wave has been presented already, however.  For example, in KFe$_2$Se$_2$,
ARPES measurements failed to find any anisotropy of the gap on the tiny $Z$-centered hole Fermi surface pockets in that system\cite{DLFengKFe2Se2}.  But if we account for {\it these} states, the appropriate model is then not (I)B but (II)A, with a 3D incipient band, which we have shown leads to a substantially enhanced $T_c$ and large gap on the incipient band.   Thus  from our perspective, these systems could still be $d$-wave or incipient  $s_\pm$, depending on details.

 In this work, we have used very simple models to investigate the fundamental nature of SC in connection with an incipient band.
We believe these models account for  most qualitative effects in the systems discussed above.
Improvement to these models can be made by including dynamical effects in the
interaction, and by including the intraband Coulomb interaction.
We note that the renormalized Coulomb pseudopotential may be effectively reduced
by an incipient band and thus
give rise to a bootstrap mechanism even if $T_c^{ph}$ in the absence of the incipient band were zero.
Finally, we have checked that one can arrive at similar conclusions to those contained in this work
 in an Eliashberg approach where the bands are
parabolic and the interaction is constant up to the Matsubara summation cutoff, similar to the BCS "box" interaction \cite{CarbottePropertiesOfBosonExchangeSC1990}.
%

Note finally that we have assumed in the numerical evaluations of the theory above that spin fluctuations with the incipient band can be significant, and in particular for case (II)A that they can be of the same order as the Fermi surface spin fluctuation interband interaction.  While this appears to us to be quite reasonable, given that magnetic interactions are defined over large energy scales of order $\Lambda_{sf}\gg E_g$,  these assumptions should be justified by concrete calculations, which are currently  in progress.

\section{Conclusion}\label{Sec:Conclusion}
We have investigated pairing on bands away from the Fermi surface within a weak-coupling multiband BCS approximation. This is possible because the pairing interaction has a finite spread around the Fermi surface. Taking advantage of this spread we identify 4 instances for a hole band: (1) Regular hole band: when the extremum of the band is far from the cutoffs; (2) Shallow band: when the extremum of a band is above the Fermi level but within the cutoff; (3) Incipient band: when the extremum of a band is  below the Fermi-surface but still within the cutoff; and (4) Vegetable band: does not take part in pairing. This article focusses on the shallow and incipient band pairing. We have further identified two cases of pairing which have qualitatively different results in the shallow and incipient regions: Case(I) where pairing is driven by interactions with the incipient band and case(II) where pairing is induced on the incipient band. We have argued that most work in the literature has only addressed case(I) and prematurely concluded that weak-coupling theories cannot be applied to certain family of FeSCs where evidence for robust pairing was found in the incipient bands. We argue in our work that case(II) has all the experimentally observed features within weak-coupling.

Our case by case results are the following: Case(I) - We have considered simple models for phonon-driven and spin fluctuation driven SC and confirmed the previously known results that pairing in the incipient scenario is strongly suppressed.  A minimum attractive strength for the SC instability is only needed in case(I)A. Case(II) - We  consider phonon-driven and spin fluctuation driven SC (from regular bands) and show that the strength of induced pairing in the shallow and incipient bands can be large, and  comparable to the pre-existing bands. The $T_c$ is enhanced quite generally in the presence of an incipient band  connected to the Fermi surface by finite-$\bf q$ spin fluctuation scattering. In this context, we discussed the bootstrapping effect of spin fluctuations on the electron-phonon SC.  All these effects  in case (II) are more pronounced in 2D,  but not qualitatively so.  We have shown that the dimensionality of the incipient band only plays a significant role for the  case(I)A model.  We have presented a simple model to study the effect of different cutoffs for the phonon-driven and spin fluctuation driven SC and indicated that the the phonon mechanisms aids the spin fluctuation mechanism.  Finally, we  discussed the results in the concrete  context of LiFeAs, FeSe intercalates and FeSe monolayers on STO, which have been reported to have similar Fermi surface missing $\Gamma$-centered hole pockets, and concluded that induced superconductivity in incipient bands may play a role in all these systems, for somewhat different reasons.

\emph{Acknowledgements}: We thank Y. Wang for useful discussions. SM is a Dirac Post-Doctoral Fellow at the National High Magnetic Field Laboratory, which is supported by the National Science Foundation via Cooperative agreement No. DMR-1157490, the State of Florida, and the U.S. Department of Energy. XC, AL and PJH were supported in part by DOE DE-FG02-05ER46236.

\bibliographystyle{apsrev4-1}
\bibliography{references}

\appendix

\section{3D incipient hole band}
\label{sec:3DIncipientHole}
In this Appendix, we derive the effect on $T_c$ and the gaps on the various bands if the DOS is not constant
as in 2D but shows the well known square root behaviour of a 3D electron
gas. We  give details for the single incipient band solution because
the results can be used later in the multi band models. The equation
that determines $T_{c}$ reads in this case
\begin{equation}
\frac{1}{v_{{\scriptscriptstyle {\rm 3D}}}}=-\int_{-\Lambda}^{E_{{\rm g}}}{\rm d}\varepsilon\rm{Re}\sqrt{\frac{E_{{\rm g}}-\varepsilon}{\vert E_{{\rm g}}\vert}}\frac{1}{2\varepsilon}{\rm tanh}(\frac{\varepsilon}{2T_{c}})\,,\label{eq:OneBand3DIncipient}
\end{equation}
where $v_{{\scriptscriptstyle {\rm 3D}}}=-\sqrt{\vert E_{{\rm g}}\vert}a\vert V_{{\rm ph}}\vert$
and $a=(2m)^{\frac{3}{4}}/(2\pi^{2})$. Thus, we are lacking
a natural reference since any density of states variation is usually
disregarded in conventional methods of SC, with the notable exception of the
density functional theory of superconductors. In the present situation
for the single incipient band we arbitrarily measure our $T_{c}$
in units of the value at $E_{g}=\Lambda$.
The coupling at our reference $E_{g}=\Lambda$ is then simply $v_{{\rm 3D}}^{0}=-a\vert V_{{\rm ph}}\vert\sqrt{\Lambda}$.
Before we resort to numerics, again, we want to discuss special cases.

\begin{figure}[htbp]
\begin{centering}
\includegraphics[width=0.8\columnwidth]{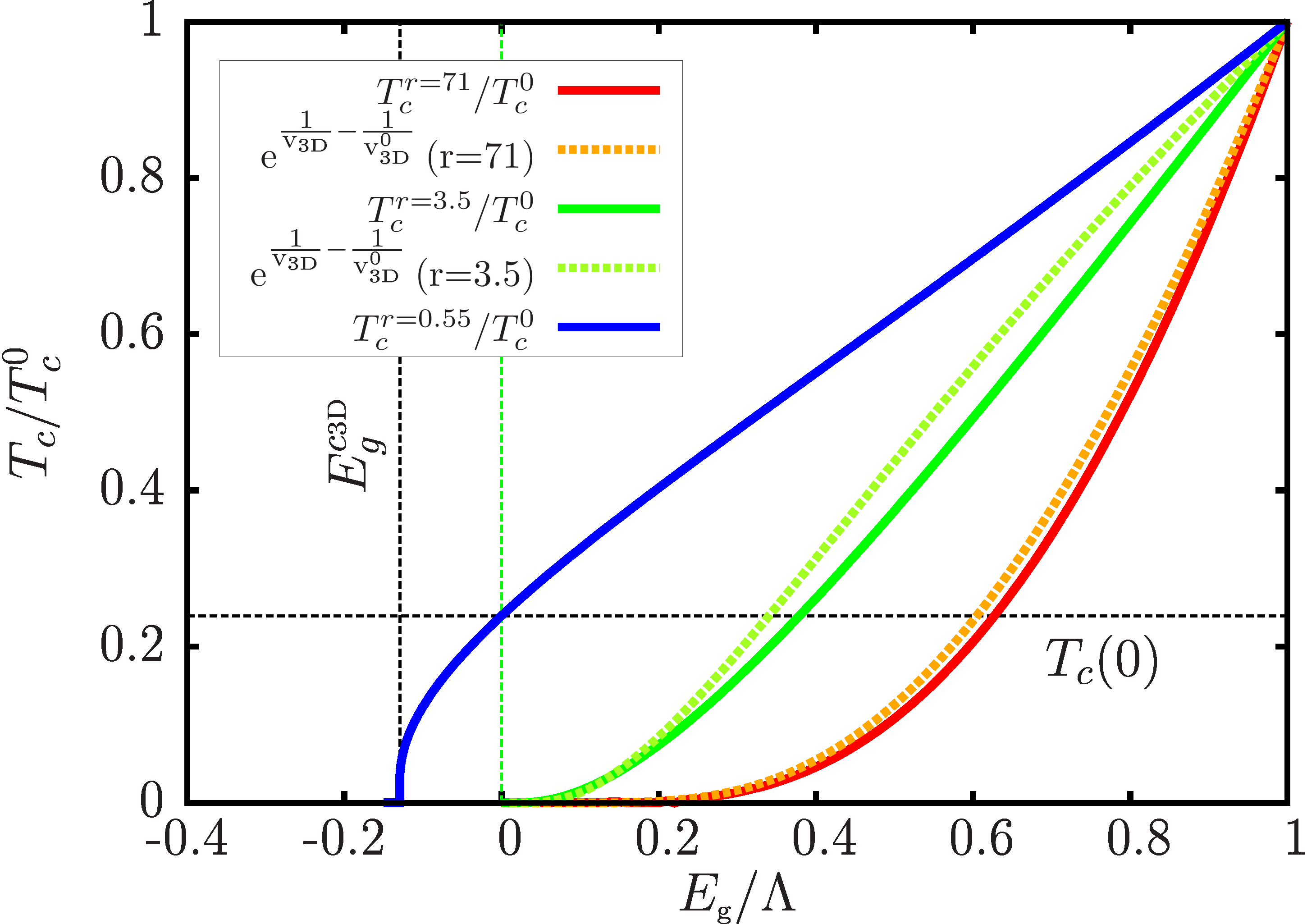}
\par\end{centering}
\caption{(color online) $T_{c}/T_{c}^{0}$ for a 3D hole band for $r\equiv\frac{\Lambda}{2T_{c}^{0}}=71$,
(red) $2.5$ (green) and $0.55$ (blue). The weak
coupling ratio $r=\frac{\Lambda}{2T_{c}^{0}}$ is indicated for each
data set in the legend.
The parameters for $r=71,2.5$ and $0.55$ are $v_{ph}=-0.2\sqrt{2}$,$-0.5\sqrt{2}$ and $-2.0\sqrt{2}$, respectively,
and the 2D vs 3D DOS ratio is chosen such that $m/(2\pi N_{h}^{{\rm {\scriptscriptstyle 3D}}})=1$.
We show Eq.~\eqref{eq:TcForLareEg} as a dashed line for $r=71$ and $r=3.5$ in dashed orange and dashed green,
respectively.\label{fig:TcVariation3DOneBandIncipient}}
\end{figure}

\subsection{Shallow band}\label{subsec:appendix-shallow}
In 3D, if the integral (\ref{eq:OneBand3DIncipient})  includes $\varepsilon=0$, we may introduce
$L_{0}^{{\rm {\scriptscriptstyle 3D}}}$ by
\begin{align}
L_{0}^{{\rm {\scriptscriptstyle 3D}}} & =\int_{0}^{E_{{\rm g}}}{\rm d}\varepsilon\sqrt{1-\frac{\varepsilon}{E_{{\rm g}}}}\frac{1}{2\varepsilon}{\rm tanh}(\frac{\varepsilon}{2T_{c}})+\nonumber \\
 & +\int_{0}^{\Lambda}{\rm d}\varepsilon\sqrt{1+\frac{\varepsilon}{E_{{\rm g}}}}\frac{1}{2\varepsilon}{\rm tanh}(\frac{\varepsilon}{2T_{c}})\,.\label{eq:Integral3DL}
\end{align}
In the weak coupling limit we can split the integral around a cutoff
 $E_g>C\gg T_{c}$ .
Then, Eq.~\eqref{eq:OneBand3DIncipient} reads
\begin{align}
-\frac{1}{v_{{\scriptscriptstyle {\rm 3D}}}} & =\int_{0}^{C}{\rm d}\varepsilon\frac{1}{2\varepsilon}(\sqrt{1-\frac{\varepsilon}{E_{{\rm g}}}}+\sqrt{1+\frac{\varepsilon}{E_{{\rm g}}}}){\rm tanh}(\frac{\varepsilon}{2T_{c}})+\nonumber \\
 & +\int_{C}^{E_{{\rm g}}}{\rm d}\varepsilon\frac{\sqrt{1-\frac{\varepsilon}{E_{{\rm g}}}}}{2\varepsilon}+\int_{C}^{\Lambda}{\rm d}\varepsilon\frac{\sqrt{1+\frac{\varepsilon}{E_{{\rm g}}}}}{2\varepsilon}\,.\label{eq:OneBandLowTEq}
\end{align}
If, in addition, $E_{{\rm g}}\gg C\gg T_{c}$, we obtain in the first
term an integrand proportional to ${\rm tanh}(\varepsilon/2T_{c})/\varepsilon$
and, thus, the original BCS problem except for some high energy renormalization
prefactor $P$
\begin{equation}
\label{eq:PDefinition}
\ln P=\int_{C}^{E_{g}}{\rm d}\varepsilon\frac{\sqrt{1-\frac{\varepsilon}{E_{g}}}}{2\varepsilon}+\int_{C}^{\Lambda}{\rm d}\varepsilon\frac{\sqrt{1+\frac{\varepsilon}{E_{{\rm g}}}}}{2\varepsilon}-\ln\frac{\Lambda}{C}\,,
\end{equation}
and we arrive at the solution
\begin{equation}
T_{c}^{{\rm {\scriptscriptstyle 3D}}}(E_{g}\gg T_{c})=P\frac{2{\rm e}^{\gamma}}{\pi}\Lambda{\rm e}^{\frac{1}{v_{{\scriptscriptstyle {\rm 3D}}}}}\,.
\end{equation}
Because of the fact that $E_{g}\gg C$, we see that the lower limit of the integral in $P$ of Eq.~\eqref{eq:PDefinition} will roughly cancle the $\ln(\Lambda/C)$ and, thus, $P$ is independent on $C$.
Furthermore, we observe in Fig.~\ref{fig:TcVariation3DOneBandIncipient} that for the weak coupling limit, $P(E_g)$ is constant and we may approximately write
\begin{equation}
T_{c}^{{\rm {\scriptscriptstyle 3D}}}(E_{{\rm g}}\gg T_{c})\approx T_{c}^{0}{\rm e}^{\frac{1}{v_{{\scriptscriptstyle {\rm 3D}}}}-\frac{1}{v^{0}_{{\scriptscriptstyle {\rm 3D}}}}}\,.\label{eq:TcForLareEg}
\end{equation}
This analysis is always valid if $T_{c}$ is small enough. We conclude
for $E_{{\rm g}}\gg T_{c}$ that superconductivity in a 3D free electron
band is induced by an arbitrary small attractive interaction via the
BCS essential singularity in the weak coupling limit. Due to the dependence
on the DOS, the effective coupling changes with $E_{g}$ as $\sim\sqrt{E_{g}}$.

\subsection{Lifshitz transition}\label{subsec:appendix-Lifshitz}

In the following, we show that even for $E_{{\rm g}}=0$ already,
we can find parameters, such that $T_{c}$ vanishes and, thus, the
simple BCS picture is substantially modified. Putting $E_{{\rm g}}=0$,
we find
\begin{equation}
\frac{\sqrt{\Lambda}}{v_{{\rm 3D}}^{0}}=-\int_{0}^{\Lambda}{\rm d}\varepsilon\frac{{\rm tanh}(\frac{\varepsilon}{2T_{c}})}{2\sqrt{\varepsilon}}
\label{eq:3DLif}
\end{equation}
Even for $T_{c}\rightarrow0,$ the resulting equation is integrable  while the 2D analog diverges.
 Moreover, in the limit $T_{c}\rightarrow0$, the above equation requires $v_{{\rm 3D}}^{0}=1$ to be satisfied.
If the coupling gets smaller,
no choice of $T_{c}$ can make the integral  large enough to match
 $v_{{\rm 3D}}^{0}$ and only the trivial solution $T_c=0$ is possible.
Note that the integral in Eq.~\eqref{eq:3DLif} can be scaled with the result
\begin{equation}
-\frac{\sqrt{\Lambda}}{v_{{\rm 3D}}^{0}}=\sqrt{\frac{T_{c}}{2}}\int_{0}^{\frac{\Lambda}{2T_{c}}}{\rm d}\xi\frac{{\rm tanh}(\xi)}{\sqrt{\xi}}\,.
\end{equation}
Proceeding by partial integration and assuming the weak coupling limit, we find
\begin{equation}
-\frac{\sqrt{\Lambda}}{v_{{\rm 3D}}^{0}}=\sqrt{\Lambda}-\sqrt{T_{c}}\phi\,,\label{eq:3DOneBandIntegralLowEg}
\end{equation}
where
\begin{equation}
\phi=\sqrt{2}\int_{0}^{\infty}{\rm d}\xi\sqrt{\xi}{\rm sech}^{2}(\xi)=1.072\,.
\end{equation}
We solve Eq.~\eqref{eq:3DOneBandIntegralLowEg} with the result
\begin{equation}
T_{c}(E_{{\rm g}}=0)=\frac{\Lambda}{\phi^{2}}\bigl(1+{v_{{\scriptscriptstyle {\rm 3D}}}^{0}}^{-1}\bigr)^{2}\,.\label{eq:ZeroTc3D}
\end{equation}
Eq.~\eqref{eq:3DOneBandIntegralLowEg} requires $-v_{{\scriptscriptstyle {\rm 3D}}}^{0}>1$
to have a real solution for $T_{c}$ and points out that for sufficiently
low couplings, SC is completely suppressed already when the band touches
the Fermi level.

\subsection{Incipient band}

If $-E_{{\rm g}}\gg T_{c}$, on the other hand, the integral in Eq.~\eqref{eq:OneBand3DIncipient}
is only weakly dependent on $T_{c}$ since ${\rm tanh}(\frac{\varepsilon}{2T_{c}})\approx1$
and, as in 2D, we arrive at the conclusion that superconductivity
is completely suppressed. To determine $E_{{\rm g}}^{c{\rm {\scriptscriptstyle 3D}}}$,
where $T_{c}(E_{{\rm g}})$ vanishes, we solve the integral for ${\rm tanh}(\varepsilon/2T_{c})=1$
\begin{align}
\int_{\vert E_{{\rm g}}\vert}^{\Lambda}{\rm d}\varepsilon\frac{\sqrt{\varepsilon-\vert E_{{\rm g}}\vert}}{2\varepsilon} & =\sqrt{\Lambda}\Bigl(\sqrt{1-\frac{\vert E_{{\rm g}}\vert}{\Lambda}}\nonumber \\
 & -\sqrt{\frac{\vert E_{{\rm g}}\vert}{\Lambda}}{\rm ArcCos[\sqrt{\frac{\vert E_{{\rm g}}\vert}{\Lambda}}]}\Bigr)\label{eq:3DBandIntegralMuchBelowEf}
\end{align}
and expand for small $E_{{\rm g}}/\Lambda$ that we combine with Eq.~\eqref{eq:OneBand3DIncipient}
with the result
\begin{equation}
E_{{\rm g}}^{c{\rm {\scriptscriptstyle 3D}}}=2\Lambda[-1-(v_{{\scriptscriptstyle {\rm 3D}}}^{0})^{-1}+\frac{\pi^{2}}{4}-\frac{\pi}{\sqrt{2}}\sqrt{-1-(v_{{\scriptscriptstyle {\rm 3D}}}^{0})^{-1}+\frac{\pi^{2}}{8}}]\,.\label{eq:CriticalEg3D}
\end{equation}
Note that this equation determines the critical $E_{{\rm g}}$ only
if $T_{c}$ is not already zero at $E_{g}=0$. The reason is that
Eq.~\eqref{eq:3DBandIntegralMuchBelowEf} assumes that Eq.~\eqref{eq:OneBand3DIncipient}
can be satisfied for any choice of $T_{c}$ which, as noted earlier,
is not the case in a weakly coupled system. From Eq.~\eqref{eq:ZeroTc3D}
for $T_{c}(0)$ we expect that setting $v_{{\scriptscriptstyle {\rm 3D}}}^{0}=1$
in the above Eq.~\eqref{eq:CriticalEg3D} we are at the transition
and in fact $E_{{\rm g}}^{c{\rm {\scriptscriptstyle 3D}}}(v_{{\scriptscriptstyle {\rm 3D}}}^{0}=-1)=0$.
Thus, the formula Eq.~\eqref{eq:CriticalEg3D} only applies for $v_{{\scriptscriptstyle {\rm 3D}}}<-1$.
The limit for $v_{{\scriptscriptstyle {\rm 3D}}}^{0}\rightarrow-\infty$
of $E_{g}/\Lambda$ is $(-4+\pi^{2}-\pi\sqrt{\pi^{2}-8})/2\approx0.79$.
For the numerical investigation over the entire range of $E_{{\rm g}}$,
as mentioned above we arbitrarily fix our reference $T_{c}^{0}$ to
the value for $E_{{\rm g}}=\Lambda$ and evaluate the integral numerically.
The resulting $T_{c}/T_{c}^{0}$ is shown in Fig.~\ref{fig:TcVariation3DOneBandIncipient}.

\begin{figure*}[htbp]
\begin{minipage}[t]{0.333\textwidth}%
\begin{center}
\includegraphics[width=1.0\columnwidth]{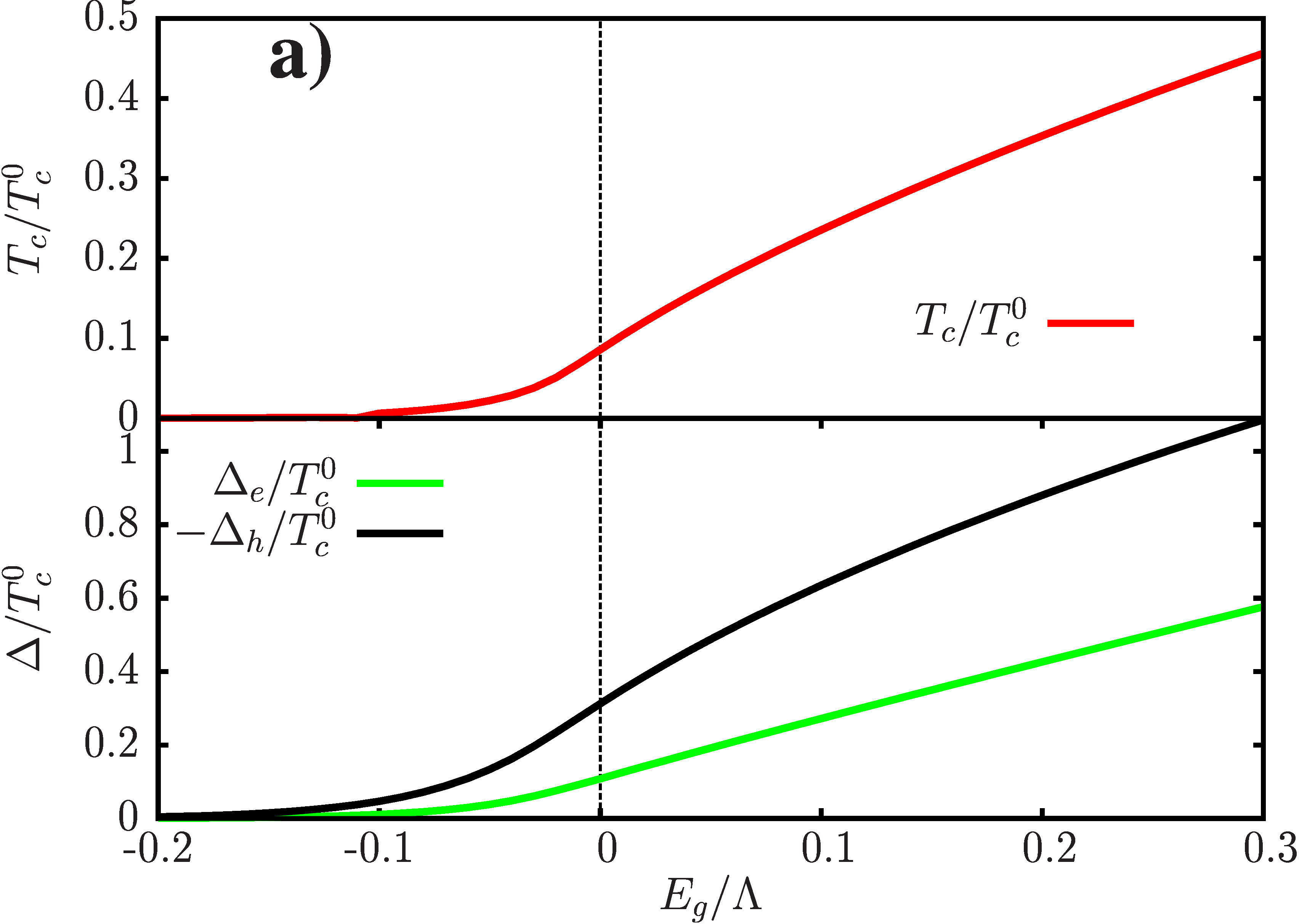}
\par\end{center}%
\end{minipage}\nolinebreak%
\begin{minipage}[t]{0.333\textwidth}%
\begin{center}
\includegraphics[width=1.0\columnwidth]{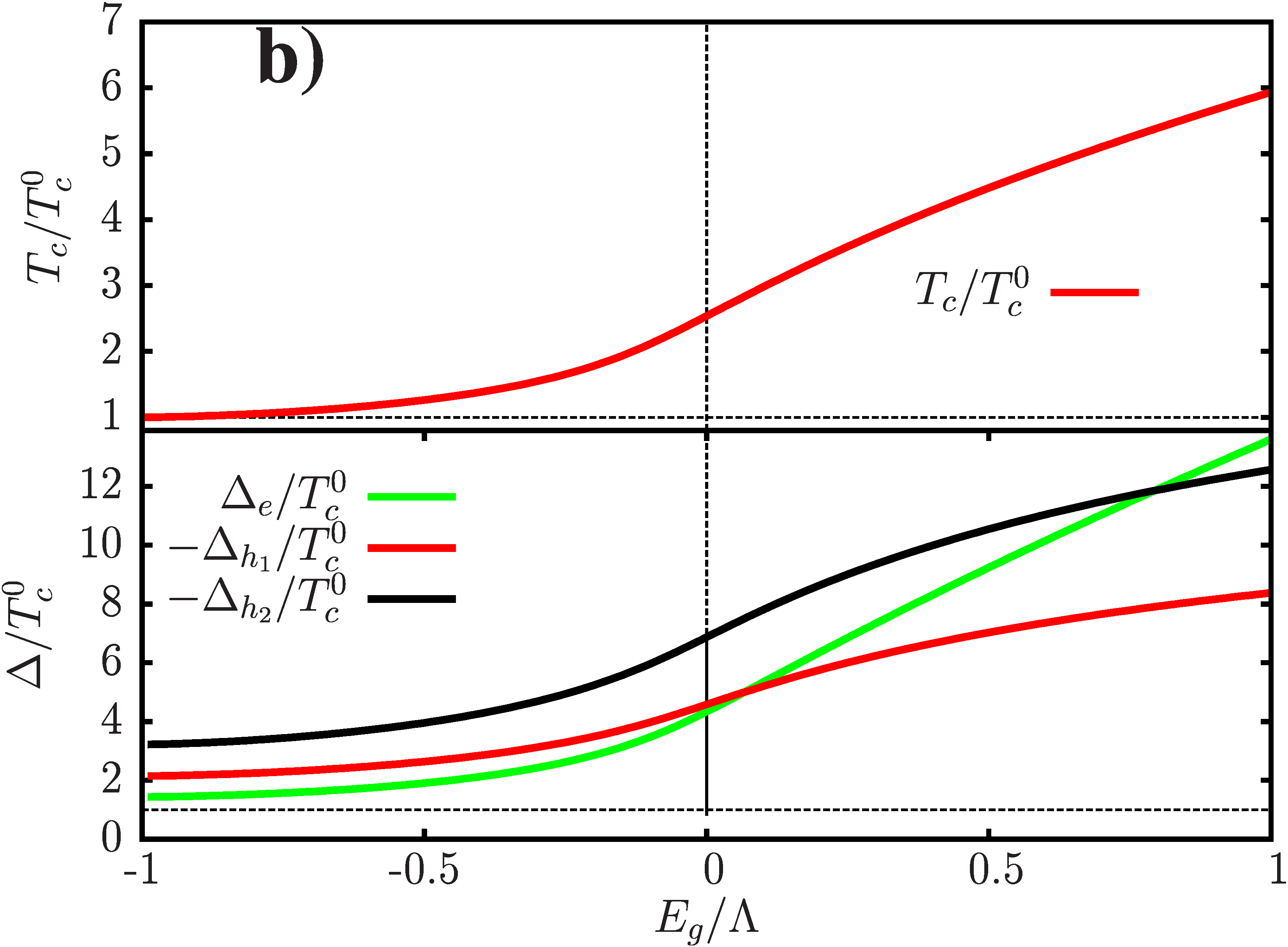}
\par\end{center}%
\end{minipage}\nolinebreak%
\begin{minipage}[t]{0.333\textwidth}%
\begin{center}
\includegraphics[width=1.0\columnwidth]{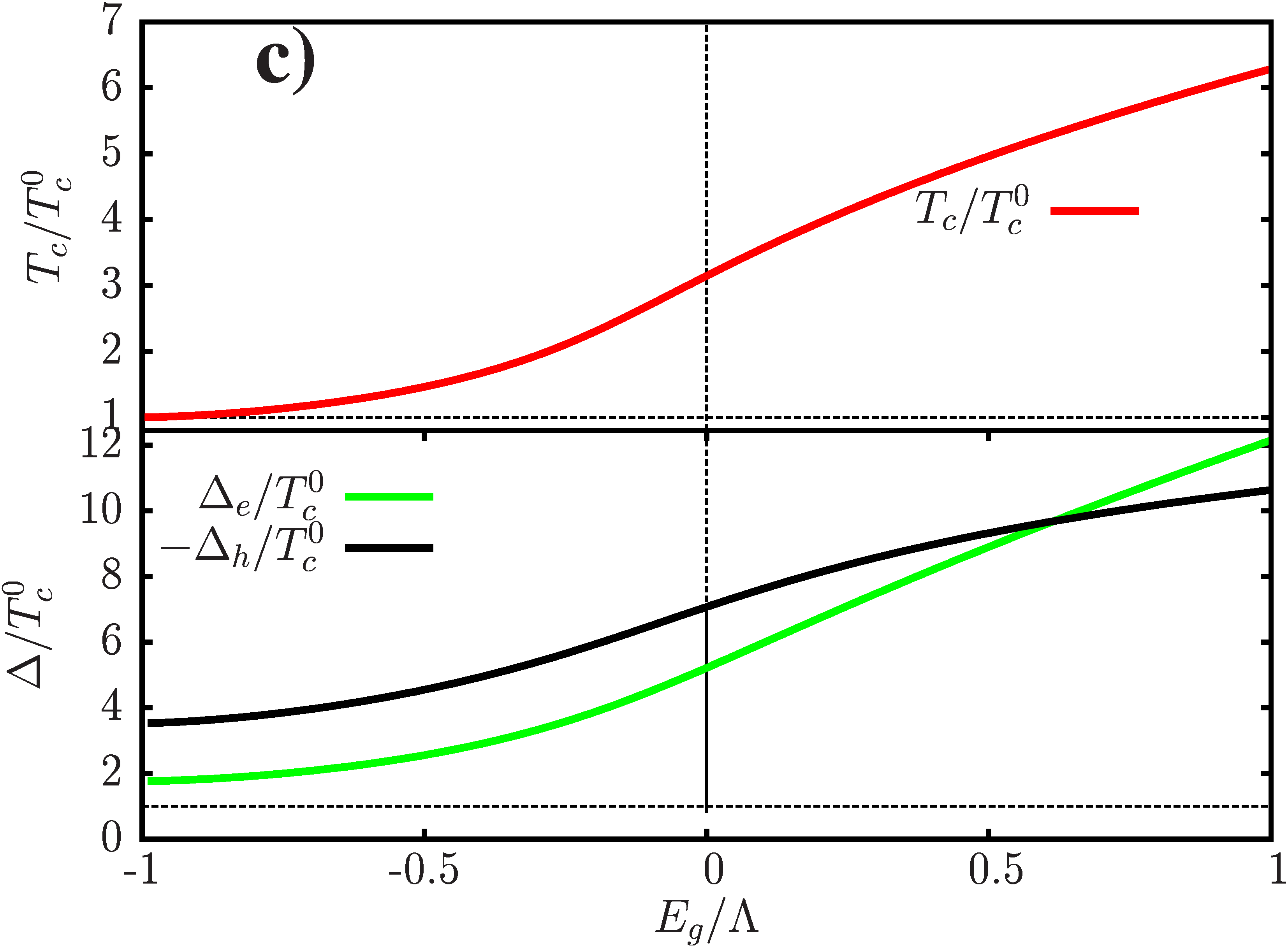}
\par\end{center}%
\end{minipage}

\caption{(color online) Comparison of 3D with 2D results for the  cases (I)B, (II)A and (II)B in panel (a), (b) and (c), respectively. We specify the DOS ratio so that $m/(2\pi N_{h}^{{\rm {\scriptscriptstyle 3D}}})=1$.
(a) The coupling parameters are $v_{sf}=-0.3$ for (a), $v_{sf_1}=v_{sf_2}=0.3$ for (b) and $v_{ph}=-v_{sf}=-0.3$ for (c). The plots (a), (b) and (c)
have to be compared to the Figs.~\ref{fig:2} (b),\ref{fig:3} and \ref{fig:4}, respectively.
\label{fig:Recollection3DIncipient}}
\end{figure*}

\subsection{Effect of a 3D incipient hole band in the cases (I)B, (II)A and (II)B}

While we have seen that an incipient 3D band requires strong coupling
to be SC at the Lifshitz transition, the conclusion that such a 3D
band does not take part in multiband SC cannot be drawn at this stage.
 As we have already discussed in the main text, in order
to understand why this is the case, consider the general difference
 between the integral of Eq.~\eqref{eq:OneBand3DIncipient} and the 2D
case of Eq.~\eqref{eq:3}.
 The additional square root lifts the BCS singularity in the integral Eq.~\eqref{eq:gaps}.
What determines
the enhancement of $T_{c}$, however, is the integral in Eq.~\eqref{eq:3DBandIntegralMuchBelowEf}.
If the incipient hole band has a 3D dispersion, we need to replace
$L_{h}$ of Eq.~\eqref{eq:def4} and $L_{h_{2}}$ of Eq.~\eqref{eq:gaps} with
$(N_{h}^{{\rm {\scriptscriptstyle 3D}}}2\pi/m)L_{0}^{{\rm {\scriptscriptstyle 3D}}}$
or $(N_{h}^{{\rm {\scriptscriptstyle 3D}}}2\pi/m)L_{0}^{{\rm {\scriptscriptstyle 3D}}}$
respectively.
We repeat the calculations for the cases (I)B, (II)A and (II)B
for a 3D incipient hole band and present the result in the Fig.~\ref{fig:Recollection3DIncipient}.

\end{document}